\documentstyle[12pt,epsf,epsfig,cite]{article}

\def\etalk{{ et al., }}
\newcommand{\scaption}[1]{\caption{\protect{\footnotesize  #1}}}

\newcommand{\av}[1]{\mbox{$ \langle #1 \rangle $}}
\def\r0{$\rho^{0}$}

\newcommand{\Wgp}{W}
\newcommand{\avWgp}{\langle\Wgp\rangle}

\newcommand{\gapprox}{\stackrel{>}{_{\sim}}}
\newcommand{\lapprox}{\stackrel{<}{_{\sim}}}
\newcommand{\pom}{\rm I\!P}
\newcommand{\reg}{\rm I\!R}
\newcommand{\pt}{p_{_T}}
\newcommand{\mx}{M_{_X}}
\newcommand{\my}{M_{_Y}}
\newcommand{\alphapom}{\alpha_{_{\rm I\!P}}}
\newcommand{\alphareg}{\alpha_{_{\rm I\!R}}}

 1
 1
 1

\def\Q2{$Q^2$}

\mathchardef\Lcur="324C

\def\qr2{{Q^2\over2}}
\def\q2{{Q^2}}

\def\be{\begin{equation}}
\def\ee{\end{equation}}
\begin{document}

%\pagestyle{myheadings}
%\markboth{}{DRAFT OF PAPER}

%\special{!userdict begin /bop-hook{gsave
%  /Times-Roman findfont 180 scalefont setfont
%  108 585 moveto 0.90 setgray 305 rotate ( draft) show
%  grestre} def end}

\begin{titlepage}
\begin{flushleft}
DESY 97--009 \hspace{10cm} ISSN 0418-9833 \\
January 1997
\end{flushleft}
\vspace*{3.cm}
\begin{center}
\begin{Large}
   \boldmath \bf{Diffraction Dissociation in \\
Photoproduction at HERA\\}
  \unboldmath

  \vspace*{2.cm} H1 Collaboration \\ 
\end{Large}

\vspace*{1cm}

\end{center}

\vspace*{1cm}

\begin{abstract}
  \noindent A study is presented of the process $\gamma p
  \rightarrow XY$, where there is a large rapidity gap between the systems $X$
  and $Y$. Measurements are made of the differential cross section as
  a function of the invariant mass $\mx$ of the system produced at the photon
  vertex.  Results are presented at centre of mass energies of $\avWgp = 187
  \ {\rm GeV}$ and $\avWgp = 231 \ {\rm GeV}$, both where the proton
  dominantly remains intact and, for the first time, where it dissociates.
  Both the centre of mass energy and the $\mx^2$ dependence of HERA data and
  those from a fixed target experiment may simultaneously be described in a
  triple-Regge model. The low mass photon dissociation process is found to be
  dominated by diffraction, though a sizable subleading contribution is
  present at larger masses.  The pomeron intercept is extracted and found to
  be $\alphapom(0) = 1.068 \pm 0.016 \ {\rm (stat.)} \ \pm 0.022 \ {\rm
    (syst.)} \ \pm 0.041 \ {\rm (model)}$, in good agreement with values
  obtained from total and elastic hadronic and photoproduction
  cross sections.  The diffractive
  contribution to the process $\gamma p \rightarrow Xp$ with $\mx^2 / \Wgp^2
  < 0.05$ is measured to be 
$22.2 \pm 0.6 \ ({\rm stat.}) \pm 2.6 \ ({\rm syst.}) \ 
  \pm 1.7 \ ({\rm model}) \, \%$ of the total $\gamma p$ cross section at
  $\av{\Wgp} = 187 \ {\rm GeV}$.
\end{abstract}
\end{titlepage}

\vfill
\clearpage
\begin{sloppypar}
%   H1AUTS  Author list by names, no. of authors  399
%           status: 05/11/96   09.30.34
 C.~Adloff$^{35}$,                %WUPP-ST                  Adloff              
 S.~Aid$^{13}$,                   %HAM2-LEFT    8/96        Aid                 
 M.~Anderson$^{23}$,              %MANC-ST  10/95           Anderson            
 V.~Andreev$^{26}$,               %LPI -PD                  Andreev             
 B.~Andrieu$^{29}$,               %ECPL-PD                  Andrieu             
 V.~Arkadov$^{36}$,               %ZEUT-ST    10/96         Arkadov             
 C.~Arndt$^{11}$,                 %DESY-ST   1/96           Arndt               
 I.~Ayyaz$^{30}$,                 %PARI-ST       5/96       Ayyaz               
 A.~Babaev$^{25}$,                %ITEP-PD                  Babaev              
 J.~B\"ahr$^{36}$
% ...
 \\
\bigskip 

\noindent
{\footnotesize{%     H1 Institutes as appearing on publications
% ...
$ ^{11}$ DESY, Hamburg, Germany$^a$ \\
% ...
$ ^{13}$ II. Institut f\"ur Experimentalphysik, Universit\"at Hamburg,
                                     Hamburg, Germany$^ a$  \\
% ...
$ ^{23}$ Physics Department, University of Manchester,
                                          Manchester, UK$^ b$\\
% ...
$ ^{25}$ Institute for Theoretical and Experimental Physics,
                                                 Moscow, Russia \\
 $ ^{26}$ Lebedev Physical Institute, Moscow, Russia$^ f$ \\
% ...
$ ^{29}$ LPNHE, Ecole Polytechnique, IN2P3-CNRS,
                             Palaiseau, France \\
 $ ^{30}$ LPNHE, Universit\'{e}s Paris VI and VII, IN2P3-CNRS,
                              Paris, France \\
% ...
$ ^{35}$ Fachbereich Physik, Bergische Universit\"at Gesamthochschule
               Wuppertal, Wuppertal, Germany$^ a$ \\
 $ ^{36}$ DESY, Institut f\"ur Hochenergiephysik,
                              Zeuthen, Germany$^ a$\\
% ...
$ ^a$ Supported by the Bundesministerium f\"ur Bildung, Wissenschaft,
        Forschung und Technologie, FRG,
        under contract numbers 6AC17P, 6AC47P, 6DO57I, 6HH17P, 6HH27I,
        6HD17I, 6HD27I, 6KI17P, 6MP17I, and 6WT87P \\
 $ ^b$ Supported by the UK Particle Physics and Astronomy Research
       Council, and formerly by the UK Science and Engineering Research
       Council \\
% ...
$ ^f$ Supported by the Deutsche Forschungsgemeinschaft \\
% ...
}}

\end{sloppypar}

%% ========================== Introduction ==========================
\section{Introduction}
\label{intro}
At the HERA electron-proton collider the bulk of the cross section
corresponds to {\em photoproduction}, in which a beam electron is
scattered through a very small angle and a quasi-real photon interacts
with the proton. For such small photon virtualities the dominant
interaction mechanism takes place via the fluctuation of the photon
into a hadronic state~\cite{sakurai:vdm} which interacts with the
proton via the strong force. High energy photoproduction therefore
exhibits broadly similar characteristics to hadron-hadron collisions,
but with cross sections typically reduced by factors of order the fine
structure constant. Whilst the regime of asymptotic freedom in strong
interactions can be described using perturbative QCD, the overwhelming
majority of the cross section in which no hard scale is present
remains far from understood.  The study of soft interactions between
photons and protons provides an excellent environment in which to
study both the hadronic manifestation of the photon and the dynamics
of the long range component of the strong force.

The dependence on centre of mass energy of elastic and, via the optical
theorem, total hadron-hadron cross sections has been remarkably well
described in a large kinematic domain by Regge
phenomenology~\cite{regge:theory}. In this framework, 
interactions take place via the $t$-channel exchange of
reggeons related to mesons~\cite{chew:frautschi}
and of the leading vacuum singularity, the
pomeron ($\pom$)~\cite{chew:pomeron}. The pomeron is
the mediator of diffractive\footnote{In this paper, the word 
`diffractive' is used synonymously with $t$-channel pomeron 
exchange; $s$-channel approaches to diffraction~are discussed in
\cite{diss:predict,schannel}.} processes. At asymptotically large energies, 
pomeron exchange dominates the elastic channel, such that both elastic and 
total cross sections display a slow increase with centre of mass
energy.  Interactions in which one
or both of the hadrons dissociates to higher mass states also
occur~\cite{diss:predict}. Such processes are characterised by the presence of
large regions of rapidity space in which no hadrons are produced
and are dominated by diffractive exchange at large
centre of mass energy $\sqrt{s}$ and small dissociation mass $M$. The
inclusive dissociation mass distribution may be treated via
Mueller's generalisation of the optical theorem~\cite{mueller:opt}, such that
an appropriate Regge description involves diagrams that contain 
three-reggeon couplings.

Experimental results on dissociative processes and their theoretical
description are extensively covered in a number of review
articles~\cite{kaidalov:review,alberi:goggi,goulianos:review,zotov:tsarev}.
At sufficiently high energy, differential dissociation cross sections 
${\rm d} \sigma / {\rm d} \mx^2$ are
approximately independent of $s$ and fall like 
$1 / M^2$~\cite{ua4:diss,ua5:diss}. The dependence 
on the 4-momentum transfer squared $t$ is approximately 
exponential at small $t$. The highest energy $p \bar{p}$
experiments~\cite{e710:diss,cdf:diss} have confirmed
this behaviour, but have reported weaker dependences of
integrated diffractive cross sections on $s$ than is
predicted by Regge models based on fits to total and elastic cross sections.
This is often taken to be the first hint of the influence of unitarity
bounds~\cite{pumplin:bound}
on diffractive cross sections and a number of mechanisms that
unitarise the cross section have been
proposed~\cite{screening,glm:screen}.

As in the case of hadron-hadron interactions, distinct elastic and dissociative
photoproduction subprocesses are usually
distinguished. Each of these may be discussed in terms of the diagram for the
process $\gamma p \rightarrow XY$ shown in figure~\ref{diffproc}. The
elastic (EL) case, $\gamma p \rightarrow Vp$, corresponds to the
situation where the system $X$ consists of a vector
meson state, $V = \rho^0$, $\omega$, $\phi$, $\ldots$, and $Y$ is a proton.
For single photon dissociation (GD), $\gamma p
\rightarrow Xp$, the proton remains intact and the photon
dissociates to a continuum of states with invariant
mass $\mx$. Single proton dissociation (PD), 
$\gamma p \rightarrow VY$, describes the case where a
vector meson is produced at the photon vertex, and the proton dissociates to
a system of mass $\my$. Finally, in the case of double dissociation
(DD), $\gamma p \rightarrow XY$, both interacting particles dissociate.

Results from HERA on the total $\gamma p$ cross section
\cite{zeus:stot,h1:stot} and exclusive light vector meson photoproduction
\cite{zeus:rho,h1:rho,zeus:omega} are well described in the Regge picture in
terms of a single pomeron trajectory, consistent with that obtained by
Donnachie and Landshoff in fits to $pp$ and $\bar{p}p$ total cross sections
\cite{dl:stot}.  Differential cross sections for the GD process have been
measured in a fixed target photoproduction experiment \cite{E612:gp}.  In
\cite{zeus:stot} overall cross sections for the EL and the sum of the GD, PD
and DD processes were measured at an average centre of mass energy $\av{\Wgp}
= 180 \ {\rm GeV}$ and in \cite{h1:stot}, the EL, GD and PD contributions
were presented as a function of the unmeasured DD cross section at $\av{\Wgp}
= 200 \ {\rm GeV}$. No measurements have yet been made at HERA that fully
unfold the PD, GD and DD components or that do not rely on model dependent
extrapolations of dissociation mass spectra.

In this paper, the mass $\mx$ is reconstructed directly in the central 
components of the H1 detector. In addition to the central detectors,
components of H1 that are sensitive to hadronic activity very close to the 
outgoing proton direction are used to constrain the mass $\my$. This
leads to measurements of ${\rm d} \sigma / {\rm d} \mx^2$, 
both for events in which the proton dominantly remains intact
(closely corresponding to the EL and GD processes) and where it dissociates
to low mass states (corresponding to the PD and DD processes). In
section~\ref{sigmadef}, the processes measured are defined at the level of
final state hadrons and kinematic variables are introduced.
Section~\ref{h1det} briefly describes the H1 detector and section~\ref{mcmods}
outlines the Monte Carlo simulations used in the measurement.  Event
selection, kinematic reconstruction methods and other details of the
experimental procedure are covered in section~\ref{expmeth}.  In
section~\ref{results} the measured cross sections are presented, the GD
process is decomposed according to a triple-Regge prescription and the
dissociation mass and centre of mass energy dependences of the diffractive
contribution are discussed.

%% ==================Measured Cross Section =====================

\section{Cross section definitions and kinematics}
\label{sigmadef}
Dissociation mass spectra have traditionally been measured by
tagging leading final state protons and inferring the mass of the
dissociation products from the 4-vectors of the beam and tagged protons.  In
this analysis a complementary approach is taken, making use of the fact that,
where a colourless exchange takes place, there is generally
an associated region in rapidity space
that contains no final state particles (a `rapidity gap'). The size of
the gap is related to
the masses of the final state systems produced at each vertex.

\begin{figure}[htb]
 \vspace{-0.5cm}
 \begin{center}
  \epsfig{file=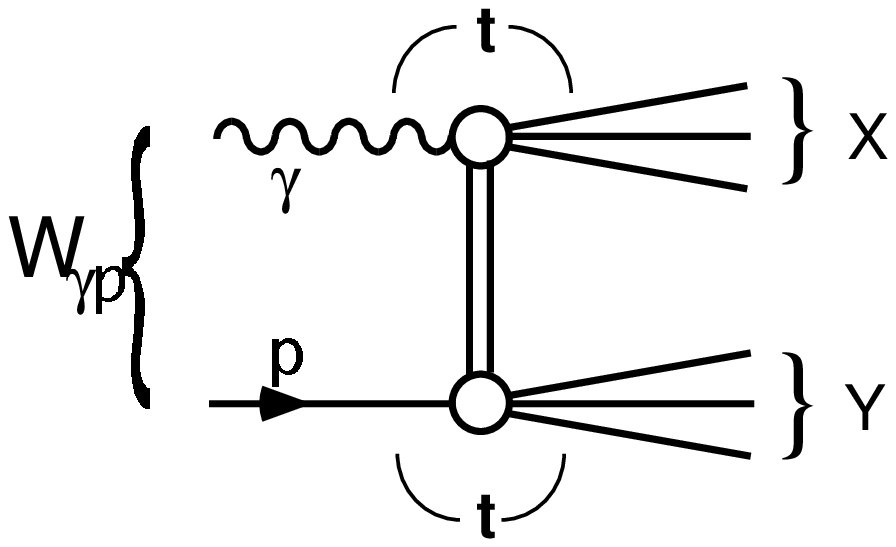,width=0.35\textwidth}
 \end{center}
 \vspace{-0.7cm}
  \scaption  {Illustration of the generic process $\gamma p \rightarrow XY$.
In the Regge pole picture, a reggeon is exchanged between the photon and the 
proton.} 
  \label{diffproc}
\end{figure}

Each event is considered in terms of the generic quasi two-body 
process, $\gamma p \rightarrow XY$, 
illustrated in figure~\ref{diffproc}. The two systems
$X$ and $Y$ are separated by the largest rapidity gap, with rapidity calculated
in the photon-proton centre of mass frame. $Y$ is the system closest to the
proton beam direction.
This scheme of event decomposition provides a means of defining hadron level
cross sections
without assumptions as to the interaction mechanism.
However, where a colourless exchange does not take place, the location of
the largest gap is determined 
by random fluctuations, the average gap size is small and
at least one of the systems $X$ or $Y$ generally has a large mass.

The kinematics of the interaction 
are specified in terms of the 4-vectors of the
incoming photon and proton, denoted by $q$ and $P$ respectively, and those
of the two distinct components of the final state, $p_{_X}$ and
$p_{_Y}$. Convenient Lorentz scalars are
\begin{eqnarray*}
\Wgp^2 = (q + P)^2 \hspace{1.5cm} t = (P - p_{_Y})^2 \hspace{1.5cm}   
\mx^2 = p_{_X}^2 \hspace{1.5cm} \my^2 = p_{_Y}^2 \ ,
\end{eqnarray*}
where $\Wgp$ is the total photon-proton centre of mass energy, $t$ is the
square of the \linebreak
4-momentum transferred and $\mx$ and $\my$ are the invariant
masses of the two final state hadronic systems. The
ratios $\mx^2 / \Wgp^2$ and $\my^2 / \Wgp^2$ most naturally
distinguish between the kinematic region where diffraction is dominant and
that in which other exchanges become important.

The mass of the system $X$ is measured using the central components of
the detector up to $\mx \simeq 30 \ {\rm GeV}$.  For $\my \gapprox 10 \ {\rm
GeV}$ the mass of the system $Y$ may also be measured. Where $\my$ is
smaller its value is constrained using the location of the rapidity gap. 
The resolution in transverse momentum is insufficient for a differential 
measurement of the $t$ dependence. Measurements of the 
differential cross section ${\rm d} \sigma / {\rm d} \mx^2$ are presented at
$\av{\Wgp} = 187 \ {\rm GeV}$ and $\av{\Wgp} = 231 \ {\rm GeV}$, for
the region $\my < 1.6 \ {\rm GeV}$ and $|t| < 1.0 \ {\rm GeV^2}$ 
and for $1.6 < \my < 15.0 \ {\rm GeV}$ and all $|t|$.

%% ========================== Detector ==========================
\section{The H1 detector}
\label{h1det}
The components of the H1 detector that are most relevant to this analysis 
are briefly described here. More detailed information can be found 
in \cite{h1:detector}. 
The coordinate system convention for the experiment defines the
forward, positive $z$ direction as being that of the proton beam. The polar
angle $\theta$ is defined relative to this axis such that pseudo-rapidity, 
$\eta = - \ln \tan \theta / 2$, is positive in the forward region.

Final state electrons are detected in the electron tagger of the luminosity
system (eTag), which is situated at $z=-33 \ {\rm m}$ and has an
acceptance such
that $Q^2 < 10^{-2} \ {\rm GeV^2}$, where $Q^2$ describes the photon
virtuality. Bremsstrahlung
photons produced at very small polar angles leave the beam-pipe
through an exit window and are detected at $z=-103 \ {\rm m}$ in the photon
arm of the luminosity system.

Minimally biased photoproduction data samples
are obtained using trigger signals from the eTag in coincidence with a
`ToF' trigger. The ToF trigger demands at least one pair of hits in the twin 
layers of plastic
scintillator comprising the Time of Flight device, during a short time period
at each bunch crossing when particles originating from $ep$ interactions in
the vertex region are expected to arrive. The backward location of the ToF
($-3.5 < \eta < -2.0$) significantly enhances the efficiency for
triggering events at low $\mx$ compared to triggers that require 
tracks in the central parts of the detector.

The system $X$ is detected and measured using tracking and calorimeter
information from the central parts
of the detector. Charged tracks are measured in the range $-1.5 <
\eta < 1.5$ \linebreak
in the two large concentric drift chambers of the central
tracker, which lies inside a $1.15 \ {\rm T}$ solenoidal field.  Surrounding
the drift chambers, the finely segmented liquid argon calorimeter (LAr)
provides smooth and hermetic coverage in the range
$-1.5 < \eta < 3.4$. The
lead-scintillator Backward Electromagnetic Calorimeter (BEMC) completes the
coverage in the backward region for $-3.5 < \eta < -1.5$. The backward
multi-wire proportional chambers (BPC) are mounted directly onto the
surface of the BEMC.

The most forward components of H1 are used to tag hadronic activity at high
pseudo-rapidity \cite{h1:f2d,h1:f2detc,jpsi:rho,h1:jpsi}. 
At all but the smallest values
of $\my$, rescattering of primary particles with the beam-pipe and the 
surrounding material yields showers
of secondary particles which are observable in these detectors,
extending their sensitivity to pseudo-rapidities 
larger than would be expected on the basis of their geometrical
acceptance alone. The Proton Remnant Tagger (PRT) consists of double layers
of scintillator surrounding the beam-pipe at $z = 24 \ {\rm m}$ and covers
the highest pseudo-rapidity region.
Three double layers of drift chambers of the Forward Muon Detector (FMD)
cover a slightly lower pseudo-rapidity range. 
These components of H1 provide high efficiency for the
identification of primary particle production in the region
$5 \lapprox \eta \lapprox 7$. A third detector component, the plug
calorimeter, is sensitive at the lower end of this range and is used to
perform cross-checks of the measurement.
The overall range of sensitivity of the H1 detector to final state
hadron production spans 11 pseudo-rapidity units, providing excellent 
discrimination between different event topologies.

%% =======================Monte Carlo ==========================
\section{Simulations}
\label{mcmods}

Two Monte Carlo models, based on the event generators PHOJET \cite{mc:phojet}
and PYTHIA \cite{mc:pythia}, are used to determine corrections to the data
for triggering and detector inefficiency, the effects of analysis cuts and
the smearing of kinematic quantities due to detector resolution. 
The description of soft photoproduction processes by both models is based on
Regge phenomenology. Both generators produce events from all four 
elastic and dissociative subprocesses described in
section~\ref{intro} as well as events in which colour is exchanged. The
generated events are passed through a full simulation of the H1 detector and
are subjected to the same analysis chain as the data.

The lowest lying vector meson states are generated in
the ratio $\rho^0 : \omega : \phi = 14 : 1.5 : 1$, \linebreak
consistent with previous 
measurements at HERA \cite{zeus:rho,h1:rho,zeus:omega} and at lower
energy \cite{bauer:review}. The differential $t$ distributions take the
peripheral form ${\rm d} \sigma_{_{\rm EL}} / {\rm d} t \propto
e^{b_{_{_{\rm EL}}}t}$, with the slope parameter in both models 
$b_{_{\rm EL}} \simeq 11 \ {\rm GeV^{-2}}$ for $\Wgp = 200 \ {\rm GeV}$. 
Non-resonant di-pion production is accounted for by reweighting
the mass distributions produced by the models such that the $\rho^0$ line 
shape matches the
parameterisation of Ross and Stodolsky~\cite{ross:stodol}, with parameters
fixed according to previous HERA measurements \cite{zeus:rho,h1:rho}.

Dissociative events are produced in the regions $\mx^2 / \Wgp^2 < 0.1$ and
$\my^2 / \Wgp^2 < 0.1$, using
approximations to triple-pomeron predictions
with couplings constrained by low energy data. The $t$-slope 
parameters, $b_{_{\rm GD}}$, $b_{_{\rm PD}}$ and $b_{_{\rm DD}}$ 
have logarithmic
dependences on $\mx$, $\my$ and $\Wgp$.  The PD and GD slope parameters are
approximately half those for the EL reaction, and the DD slope
parameter is reduced by a further factor of approximately 2.

The PYTHIA model of dissociation processes assumes a pomeron intercept of
unity, such that the differential cross section for the GD process at fixed
$\Wgp$ takes the approximate form
\begin{eqnarray}
\frac{{\rm d} \sigma_{_{\rm GD}}}{{\rm d}t \, {\rm d} \mx^2} \sim 
\frac{1}{\mx^2} \ e^{b_{_{\rm GD}}(\mx) t} \ ,
\end{eqnarray}
with similar expressions for the differential PD and DD cross
sections. Additional factors are applied
in order to modify the distributions in kinematic regions in which a
triple-pomeron model is known to be inappropriate. Their main effects are to 
enhance the low mass components of the dissociation spectra, suppress the
production of very large masses and, in the DD case, to
reduce the probability of the systems $X$ and $Y$ overlapping in rapidity
space~\cite{mc:pythia,schuler:sjostrand}. 
GD, PD and DD states produced with $\mx
\lapprox 1.8 \ {\rm GeV}$ or $\my \lapprox 1.8 \ {\rm GeV}$ decay
isotropically to two-body final states. The fragmentation of higher mass
dissociation systems is treated in a string model, with final state hadrons 
distributed in a longitudinal phase space with limited 
transverse momentum $\pt$~\cite{jetset}.

In the PHOJET model, the first moment Finite Mass Sum Rule (FMSR)
\cite{sanda:fmsr} is applied in order to achieve a smooth transition between
the elastic, resonance and high mass continuum regions. At fixed $\Wgp$, the
GD spectrum follows the parameterisation
\begin{eqnarray}
  \frac{{\rm d} \sigma_{_{\rm GD}}}{{\rm d}t \, {\rm d} \nu} 
\sim \left( \frac{1}{\nu} \right)^{\alphapom (0)} e^{b_{_{\rm GD}}(\nu)t} \ ,
  \label{fmsr}
\end{eqnarray}
where the pomeron intercept is taken to be $\alphapom(0) = 1.07$, $\nu =
\mx^2 - M_{_0}^2 -t$ and the
`mass' $M_{_0}$ of the incoming photon is taken to be that of the $\rho$
meson.  A similar para\-meterisation was found to work well for fixed target
photoproduction data~\cite{E612:gp}. PD and DD distributions are produced in
a similar manner with the proton mass defining $M_{_0}$ for the system $Y$.
Dissociation to states in the resonance region ($\mx^2 < 5 \ {\rm GeV^2}$)
are treated through the extended Vector Dominance
Model \cite{greko:evdm} in a similar manner to EL processes. A single
effective resonance is assumed in the low mass GD region, normalised using
the FMSR and decaying to 2 or 3 particle final states. The slope parameter
$b_{_{\rm GD}}$
falls smoothly through the resonance region.  Higher mass final states are
again limited in $\pt$.

The mass distributions and ratios for the distinct subprocesses are iteratively
reweighted such that the differential cross sections of each Monte Carlo
simulation best describe the measured cross sections presented in 
section~\ref{pointsec}.

A third event generator, DIFFVM~\cite{diffvm}, models vector meson production 
at large $Q^2$, both where the proton dissociates and where it remains 
intact.
The simulation is based on Regge behaviour and the Vector Dominance 
Model and is described in more detail in \cite{jpsi:rho,h1:jpsi}.
Since DIFFVM is able to generate dissociative events both with limited $\pt$ 
and with isotropic decays of the dissociative system, it is used to check the 
sensitivity of the measurement to differing decay properties of the system $Y$.

%% ========================== Method ==========================
\section{Experimental method}
\label{expmeth}
\subsection{Selection of photoproduction events}
\label{select1}
The data used in this measurement were collected in two short dedicated runs
in 1994, when HERA was colliding positrons at $27.5\ {\rm GeV}$ with $820\ 
{\rm GeV}$ protons. The first run was taken with the vertex at its
nominal position of $\bar{z} = 4 \ {\rm cm}$ and the second with the
vertex shifted in the proton direction to $\bar{z} = 71 \ {\rm cm}$. The
latter configuration gives an enhanced efficiency for the
detection of the most backward going final state particles. The integrated 
luminosities of the two samples
are $24.7 \pm 0.4 \ {\rm nb^{-1}}$ and $23.8 \pm 1.3 \ {\rm nb^{-1}}$
respectively. The data samples have previously been used in measurements
of the total $\gamma p$ cross section~\cite{h1:stot} and of the cross section
for exclusive $\rho^0$ photoproduction~\cite{h1:rho}. 

Photoproduction events were triggered on the basis of a coincidence of 
signals from the ToF and eTag triggers described in section~\ref{h1det}.
These requirements have been shown to give good efficiency  for all
photoproduction subprocesses \cite{h1:stot}. 
The eTag is also used to measure the energy of the scattered electron, from
which the $\gamma p$ centre of mass energy $\Wgp$ is inferred.
Two ranges of $\Wgp$ are considered, covering the region in which
the acceptance of the eTag is highest. The first,
$164 < \Wgp < 212\ {\rm GeV}$, yields a
centre of mass energy averaged over the photon flux 
distribution~\cite{h1:stot,ww:flux} of 
$\av{\Wgp} = 187 \ {\rm GeV}$. The second range is
$212 < \Wgp < 251\ {\rm GeV}$, for which $\av{\Wgp} = 231 \ {\rm GeV}$.

Events arising from the Bremsstrahlung process, $ep \rightarrow ep \gamma$,
are removed by requiring no activity above $2 \ {\rm GeV}$ in the photon
detector. A correction of $1.8\%$ is made for the loss of signal due to the
random overlap of Bremsstrahlung events with $\gamma p$ interactions.

The principal background to the measurement arises from interactions of the
electron beam with residual gas particles or with the walls of the beam-pipe.
Proton-beam induced interactions also contribute to the background when they
occur in coincidence with a reconstructed electron in the eTag. This may
arise from electron-gas/wall collisions or from a Bremsstrahlung process in
which the photon is not detected. All of these sources of contamination are
suppressed by requiring either that the total energy reconstructed in the
calorimeters is greater than $1.5 \ {\rm GeV}$, or that a vertex is
reconstructed at a position in $z$ that lies within $30\ {\rm cm} \ (\simeq 3
\sigma)$ of the mean interaction point.  Any event in which a vertex is
reconstructed outside this region is rejected.  In the higher range of $\Wgp$,
where backgrounds are largest, a BPC hit is required in addition.  After these
cuts, the remaining level of background associated with the electron beam
alone is estimated from non-interacting electron pilot bunches and that
arising from the proton beam is determined from event samples in which
either the ToF or the eTag component of the trigger is relaxed. Both
backgrounds are subtracted statistically.
The subtraction varies between zero and 19\% in the measured region.

\subsection{Separation of the systems {\boldmath $X$} and {\boldmath $Y$}}
\label{separate}

In order to ensure that the systems $X$ and $Y$ are clearly separated, 
only those events that contain a region of at least two units 
of pseudo-rapidity in which no particles are reconstructed
are considered in the analysis.
This gap extends from the largest pseudo-rapidity reached by fragments of the 
system $X$ to the smallest pseudo-rapidity at which particles 
associated with the system $Y$ are observed.
Its presence implies that both $\mx$ and $\my$ are small compared to $\Wgp$.

Three distinct subsamples are defined. In sample A there is no activity 
above noise levels in the forward detectors (PRT and FMD) and the most 
forward part of the LAr ($\theta < 4.7^{\circ}$). \linebreak
These cuts are similar to those which have been used in previous H1
publications \linebreak \cite{h1:rho,h1:f2d,h1:f2detc,jpsi:rho,h1:jpsi}.
With this selection, the pseudo-rapidity gap spans at least
$3.2 \lapprox \eta \lapprox 7.0$
under nominal vertex conditions and $3.0 \lapprox \eta \lapprox 7.0$ for
shifted vertex data.
In sample B a signal is observed above noise levels
in the forward detectors and there is a
pseudo-rapidity gap of at least two units in the central detectors
(central tracker and LAr and BEMC calorimeters),
extending to the forward limit of acceptance of the LAr. Sample C is defined
in the same way as sample B, except that the gap in the central detectors
does not extend to the forward edge of the LAr. 
 
The data and simulated events that are shown in all subsequent figures are
required to pass one of these three selection criteria. There are
8685 such events in the nominal vertex data sample and 8184 in the shifted
vertex sample. 

\subsection{Reconstruction of {\boldmath $\mx$} and {\boldmath $\my$}}
\label{massrec}

The mass reconstruction has been optimised by combining tracking and
calorimeter information in an energy flow algorithm 
that avoids double counting. Primary
vertex constrained tracks are extrapolated into the calorimeter and energy
clusters increasingly distant from the extrapolated track are discounted. The
procedure continues either until the total energy of the excluded clusters
exceeds that of the track, under the assumption that the particle yielding the
track is a pion, or until all electromagnetic (hadronic) clusters within a
cylinder of radius $30 \ {\rm cm}$ ($50 \ {\rm cm}$) about 
the extrapolated track have
been removed.  Detector noise is potentially problematic, particularly at
small values of $\mx$. To minimise its effect, techniques have been developed
that reject low energy isolated clusters in the LAr. Any calorimeter deposit
that is not associated with a track or rejected as noise is accepted.

\begin{figure}[h] \unitlength 1mm
 \begin{center}
   \begin{picture}(160,60)
     \put(10,-5){\epsfig{file=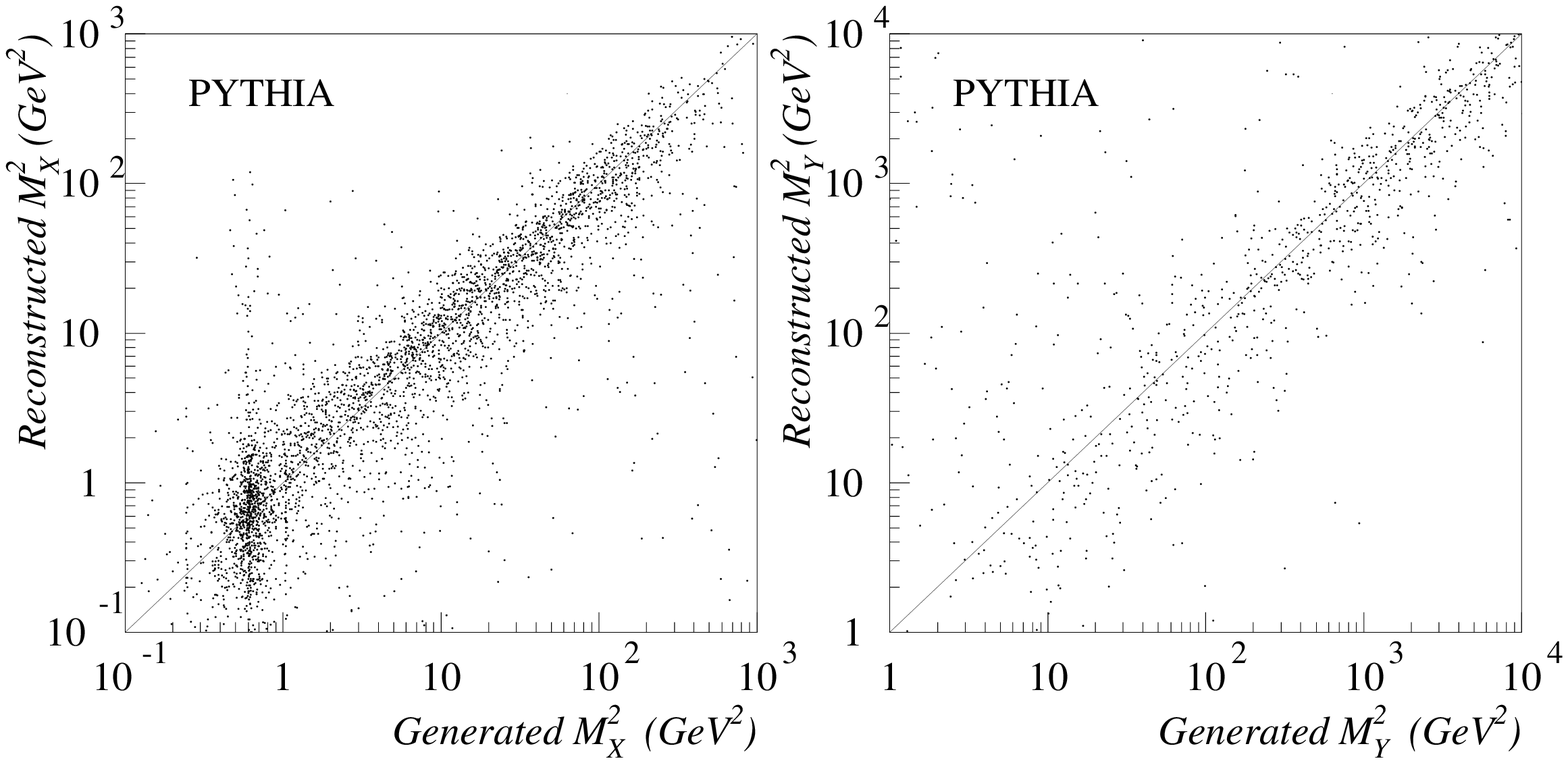,width=0.95\textwidth}}
     \put(67,15){\bf{(a)}}
     \put(137,15){\bf{(b)}}
   \end{picture}
 \end{center}
 \scaption {Correlation between reconstructed and generated values of
   (a) $\mx^2$ and (b) $\my^2$ for events in the shifted vertex PYTHIA
   simulation with $164 < \Wgp < 251 \ {\rm GeV}$.}
  \label{massdistX}
\end{figure}

The same method is used to reconstruct $\mx$ for all three samples
A, B and C. The gap requirements introduced in section~\ref{separate} 
ensure that $X$ does not extend beyond the forward acceptance limit of 
the central detector components. Losses in the backward direction 
are, however, unavoidable
both into the beam-pipe and because of the poor containment of
hadrons by the BEMC. In order to minimise the effect of such losses, the mass
of the system $X$ is reconstructed using the expression
\begin{eqnarray}
  \mx^2 = 2 \ q .p_{_X} + Q^2 + t \
\simeq \ 2 E_{\gamma} \; \sum (E+p_z)_{_X} \ \ ,
\label{eqn:mxrec}
\end{eqnarray}
where $E_{\gamma}$ is the laboratory energy of the interacting photon,
obtained from the electron energy measurement,
and $t$ and $Q^2$ have been neglected in the last expression. 
The quantity $\sum (E+p_z)_{_X}$ is summed over all hadrons reconstructed
backward of the largest pseudo-rapidity gap and has the property of being 
insensitive to the very 
backward going hadrons with $E \simeq -p_z$ that are usually not observed.
A constant correction factor of 1.07 is applied to the reconstructed $\mx$
to account for the remaining losses.
Figure~\ref{massdistX}a shows a comparison between the reconstructed and
generated values of $\mx^2$ in the PYTHIA simulation.
Good reconstruction is obtained for all masses, including those
corresponding to exclusive vector meson production.  The resolution shows a
slow improvement with increasing mass and is approximately 30\%. 

Only in sample C, where fragments from the system $Y$
are observed in the central detector components, is it possible to 
reconstruct $\my$. For this sample a similar 
approach is taken for the reconstruction of $\my$ as for $\mx$. The relation
\begin{eqnarray}
  \my^2 = 2 \ p . p_{_Y} - m_{_p}^2 + t \ \simeq \
2 E_{p} \; \sum (E-p_z)_{_Y} 
\label{eqn:myrec}
\end{eqnarray}
is used,
where $E_p$ is the energy of the proton beam and the proton mass $m_{_p}$ 
and $t$ are neglected in the last expression. Equation (\ref{eqn:myrec})
is insensitive to losses at very large
pseudo-rapidities. The remaining effect of undetected particles is accounted
for by applying a constant 
correction factor of 1.10 to the reconstructed $\my$. 
Figure~\ref{massdistX}b shows
a comparison between reconstructed and generated $\my$ values from
simulated events in which particles from the system $Y$ reach the central
detector components (sample C). 
Good reconstruction is obtained for $\my \gapprox 10 \ {\rm GeV}$.

For samples A and B the system $Y$ is not observed in the central detectors 
and direct reconstruction of $\my$ is not possible.
Instead, information is obtained from the presence or absence
of activity in the forward detectors. The location of the upper
edge of the pseudo-rapidity gap is 
correlated with $\my$ so that, with a knowledge 
of the regions of pseudo-rapidity to which the forward detectors are sensitive,
different regions of $\my$ are distinguished. 

\subsection{Selection of intervals in the invariant mass {\boldmath $\my$}}
\label{myrec}

The three subsamples defined in section~\ref{separate} are used to measure
cross sections in two ranges of $\my$. 
For sample A the system $Y$ passes unobserved down the
forward beam pipe at $\eta \gapprox 7$. 
Studies using the Monte Carlo
simulations have shown that this selection is almost 100\% efficient for
events in which $Y$ is a single proton and falls to approximately 50\% at 
$\my = 1.6 \ {\rm GeV}$. Sample A is therefore
used to measure the cross section with $\my < 1.6 \ {\rm GeV}$.
This region is likely to be dominated by the
subprocesses GD and EL in which $Y$ is a proton. 
There are also likely to be contributions
from events in which $Y$ is a different low mass baryonic system. Since neither
the isospin nor the charge of $Y$ is determined, this may be any of $n$,
$N^{\star}$, $\Delta$, $\ldots$, or a non-resonant system.

Samples B and C are used together to measure the cross section for events of 
the types DD and PD integrated over the range
$1.6 < \my < 15.0 \ {\rm GeV}$.
For sample B, $\my \lapprox 5 \ {\rm GeV}$, while 
in sample C $\my$ is usually larger. The full sample B is used along
with events from sample C for which $\my$, reconstructed using
equation (\ref{eqn:myrec}), is less than $15.0 \ {\rm GeV}$.
The restrictions imposed by the pseudo-rapidity 
gap criteria mean that the maximum 
accessible value of $\mx$ decreases as $\my$ increases. The upper
limit of $\my = 15.0 \ {\rm GeV}$ is chosen so that the measurement can be 
made over a large range of $\mx$, whilst ensuring that the upper limit 
in $\my$ lies in the region where the reconstruction is good.

\begin{figure}[h] \unitlength 1mm
 \begin{center}
  \begin{picture}(155,40)
   \put(0,-5){\epsfig{file=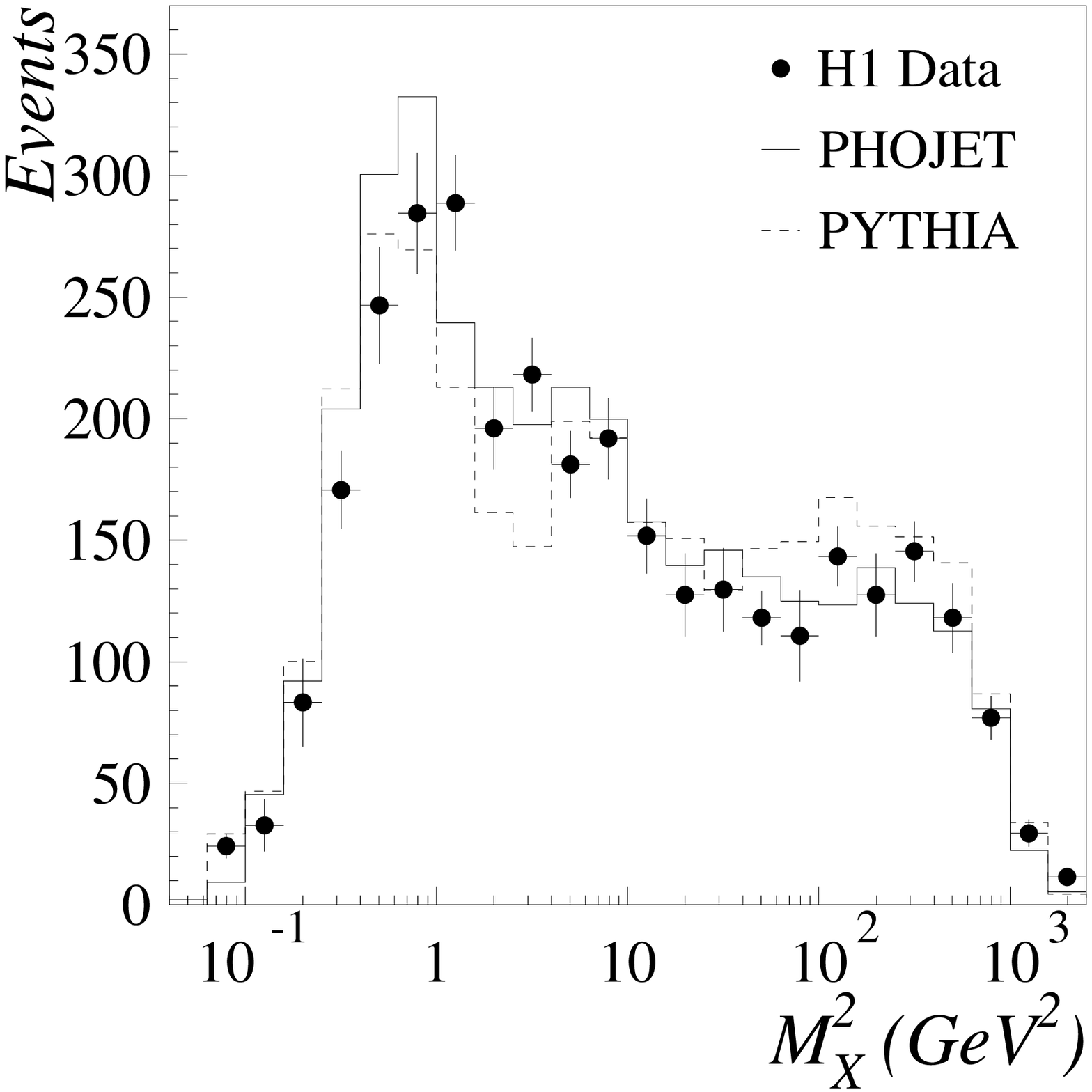,width=0.32\textwidth}}
   \put(52,-5){\epsfig{file=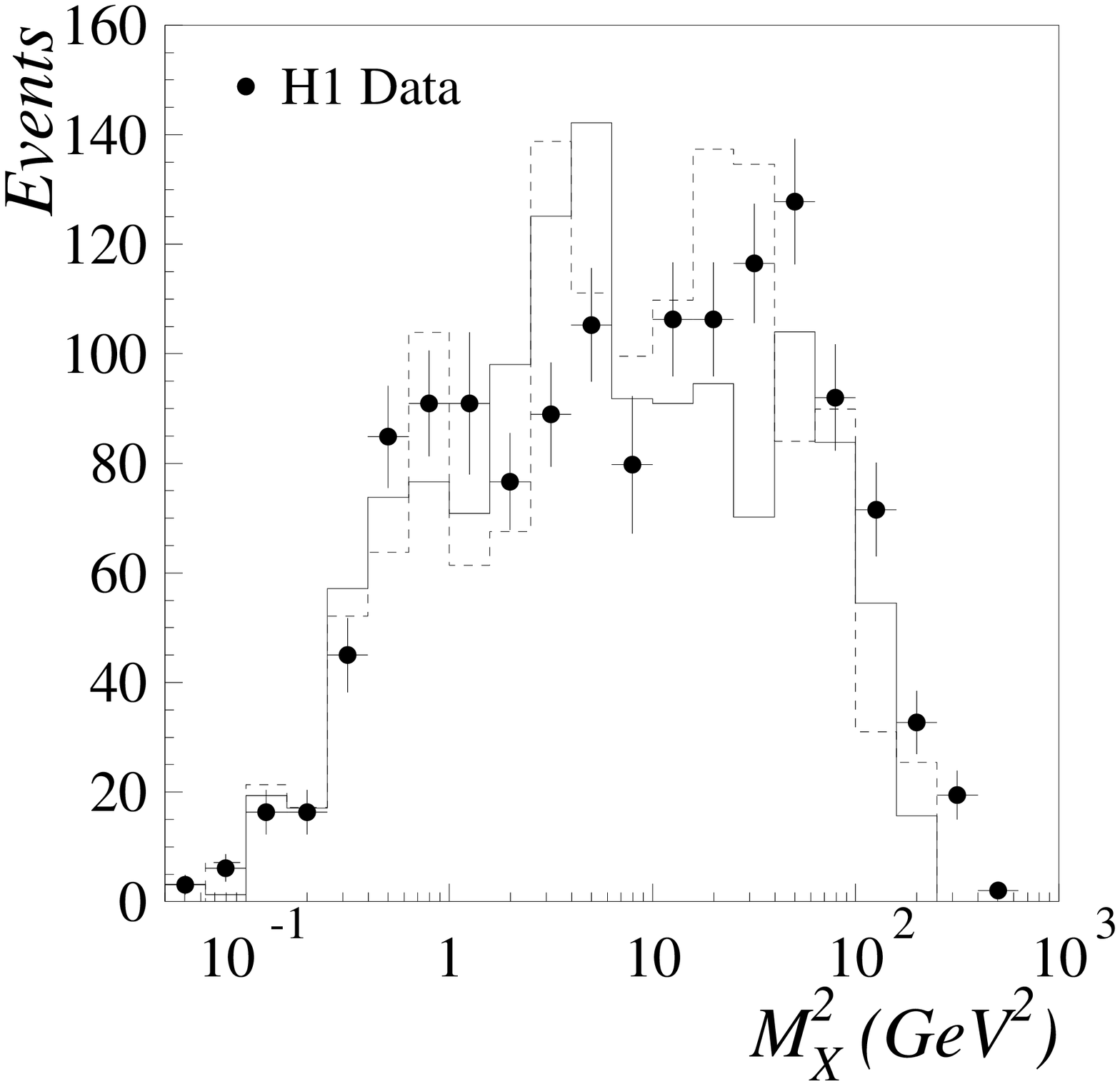,width=0.32\textwidth}}
   \put(104,-5){\epsfig{file=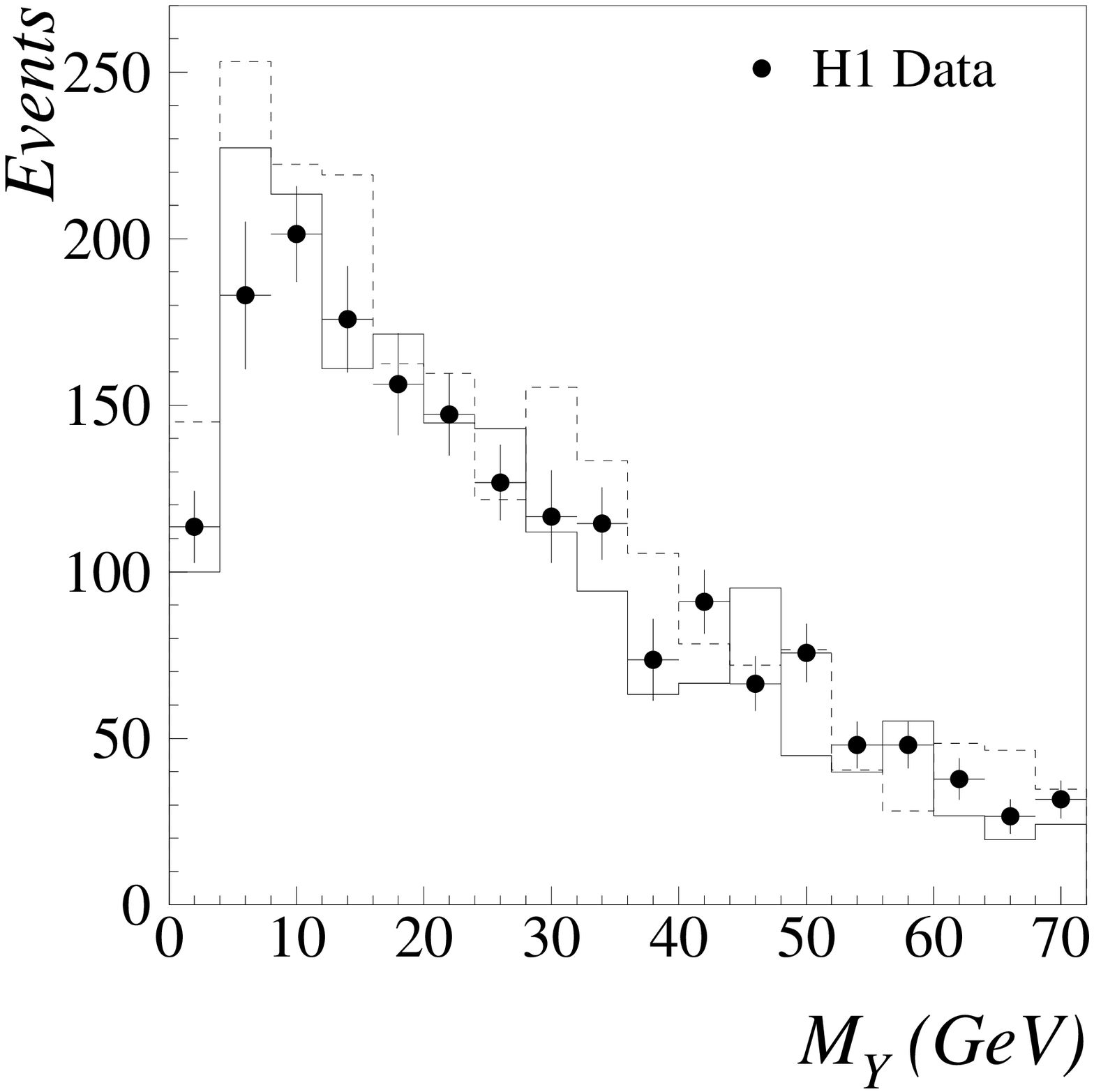,width=0.32\textwidth}}
   \put(15,6){\bf{(a)}}
   \put(68,6){\bf{(b)}}
   \put(114,6){\bf{(c)}}
  \end{picture}
 \end{center}
 \vspace{-0.2cm}
  \scaption  {Uncorrected mass distributions for the shifted 
vertex data sample in the range \protect \linebreak \protect 
$164 < \Wgp < 212 \ {\rm GeV}$ compared to 
the two simulations normalised via the total 
$\gamma p$ cross section\protect\cite{h1:stot}\protect.
(a) $\mx^2$ distribution of events from sample A, used to measure the 
cross section for the region $\my < 1.6 \ {\rm GeV}$.
(b) As in (a), but for events in samples B and C that are used to measure
the cross section for $1.6 < \my < 15.0 \ {\rm GeV}$.
(c) Distribution in $\my$ of events in sample C, where activity from $Y$
is observed in the central detector components.} 
  \label{control}
\end{figure}

Figures~\ref{control}a and~\ref{control}b show uncorrected $\mx^2$ 
distributions
for events used to make measurements in the two $\my$ intervals compared to 
the simulations. Satisfactory agreement between the data and the Monte  
Carlo models is obtained
in both cases. An elastic peak is observed in sample A, along
with a higher mass continuum that extends until the system $X$ reaches the
limit set by demanding no activity in the most forward region of the LAr.
In samples B and C no significant enhancement in the $\log \mx^2$ 
distribution is observed in
the region of the light vector mesons, and the
requirement of a gap of at least two pseudo-rapidity units in the central
detector components restricts the data to smaller values of $\mx$.
Figure~\ref{control}c shows the reconstructed distribution in proton
dissociation mass of events from data sample C.

\subsection{Extraction of cross sections}

The criteria described in section~\ref{myrec} are used to measure
cross sections as defined in section~\ref{sigmadef} for the regions 
$\my < 1.6 \ {\rm GeV}$ and $1.6 < \my < 15.0 \ {\rm GeV}$. 
The events in each $\my$ range are divided into intervals in 
$\mx^2$ and the residual contribution from beam induced background is 
subtracted in each interval using the procedure described in 
section~\ref{select1}. 

An average of the PHOJET and PYTHIA models is used to correct for
losses and migration of events in $\mx$ and $\my$.
For $|t| \gapprox 1.0 \ {\rm GeV^2}$, elastically scattered 
protons are observed in the PRT, resulting in their classification
in the wrong $\my$ interval. The simulations are therefore also used to
correct the measurement for $\my < 1.6 \ {\rm GeV}$ to the region 
$|t| < 1.0 \ {\rm GeV^2}$
and that for $1.6 < \my < 15.0 \ {\rm GeV}$ to all $|t|$.
  
The Monte Carlo models do not explicitly simulate the states $Y$ 
that may be present in the data but have different isospin or charge from
the proton.
However, the acceptance for such states is likely to be similar to that for
final states arising from diffractive
exchange and, after the reweighting procedure described in 
section~\ref{mcmods}, no additional corrections are made. A small correction
is applied for the loss of events contributing to the cross section at large
$\my$ and large $\mx$ for which the largest gap in pseudo-rapidity is smaller
than two units. The correction
necessary in unfolding from the experimental separation of the systems $X$
and $Y$ using the largest gap in pseudo-rapidity to the cross section defined
in section~\ref{sigmadef} in terms of rapidity is found to be negligible.

The mean acceptance of the eTag is $64 \pm 2 \%$ for the 
$\av{\Wgp} = 187 \ {\rm GeV}$ sample and $35 \pm 2 \%$ for 
$\av{\Wgp} = 231 \ {\rm GeV}$. The
combined acceptance of the ToF component of the
trigger and the off-line cuts
is greater than $40 \%$ for all measurements and
purities\footnote{Purity is defined as the fraction of all simulated events
reconstructed in a mass interval that are also
generated in that interval.}
are better than $30\%$ for all measured data points. The maximum values of 
$\mx^2 / \Wgp^2$ and $\my^2 / \Wgp^2$ considered are 0.041 and 0.0084 
respectively. Migration from the regions $\mx^2 / \Wgp^2 > 0.1$ and 
$\my^2 / \Wgp^2 > 0.1$ is less than $10 \%$ for all quoted measurements. 

A $\gamma p$
cross section is obtained assuming a photon flux derived from the
Weizs\"{a}cker-Williams approximation~\cite{ww:flux,h1:stot}. 
Radiative corrections
were found to be at the level of 1\% in~\cite{h1:stot} and are neglected
here. The measurements obtained from the shifted and nominal vertex data
samples are compatible within statistical errors. The two
measurements are averaged with weights determined from their statistical
errors to produce the quoted cross sections.
All measurements are presented at the bin centres in $\log \mx^2$. With
the exception of the lowest two $\mx$ intervals, where
exclusive vector meson production gives rise to a resonance 
structure~\cite{h1:rho}, 
bin centre corrections at the level of
$2 \%$ are applied, based on the fits presented in sections~\ref{fitsec} 
and~\ref{ddsec}.

For $\av{\Wgp} = 231 \ {\rm GeV}$ the
boost of the $\gamma p$ centre of mass frame relative to the laboratory is
reduced compared to that at $\av{\Wgp} = 187 \ {\rm GeV}$, such that 
measurements in the lowest $\mx$ region are no longer
possible, but larger $\mx$ values become accessible. Statistical
errors are largest at high $\Wgp$, since both the photon
flux distribution and the acceptance of the eTag
fall with increasing photon energy.

A cross check has been performed with a sample of data triggered by
a track in the central chambers~\cite{h1:stot}, for which the acceptance
becomes high at large $\mx$.  The results were consistent with those obtained
from the ToF triggered samples. The sensitivity of the results to variations
in the pseudo-rapidity gap selection criteria and mass reconstruction
methods have been checked in several
ways. No significant changes in the measured cross sections are observed when
the forward region of the LAr that must be devoid of activity in the low $\my$
measurement is enlarged to $\theta < 10^{\circ}$ or when this cut is
removed completely. There is also little change when the minimum
gap size requirement for the larger $\my$ measurement is varied between 1 and 
3 pseudo-rapidity units, when a further
forward detector, the plug calorimeter, is included, or when $\mx$ and $\my$
are reconstructed using calorimeter clusters only.

\subsection{Error analysis}
Statistical errors arise from the finite volume of data, the samples
used in the subtraction of beam induced background and the simulation samples 
used to make corrections. The
statistical error in each interval is formed by adding the contributions from
these three sources in quadrature.

Systematic errors are evaluated on a bin by bin basis.
No single contribution dominates in any region of
the measurement.
A brief summary of the systematic effects that are found to be most
important is given below.
\begin{itemize}
  \item The uncertainties in the hadronic energy scales of the BEMC and LAr
calorimeters
are $20\%$ and $5\%$ respectively. That on track measurements is $3\%$.
There is a 1\% uncertainty in the energy calibration of the eTag.
  \item The uncertainty in the efficiency of the PRT is $20\%$.
  \item A $10\%$ variation in the beam induced background, accounting
for uncertainties in the extrapolation of pilot bunch information.
  \item Uncertainties in the decomposition of the total photoproduction
cross section are accounted for with $50\%$ variations in the subprocess
ratios $\sigma_{_{\rm DD}}/\sigma_{\rm tot}^{\gamma p}$ and 
$\sigma_{_{\rm PD}}/\sigma_{\rm tot}^{\gamma p}$ in
the simulations and $20\%$ variations in the ratios
$\sigma_{_{\rm GD}}/\sigma_{\rm tot}^{\gamma p}$ and 
$\sigma_{_{\rm EL}}/\sigma_{\rm tot}^{\gamma p}$.
\item The shapes of the GD and DD mass spectra assumed in the simulations are
  individually reweighted by a factor $(\frac{1}{\mx^2})^{\pm 0.15}$. Those
  of the PD and DD processes are reweighted by a factor
  $(\frac{1}{\my^2})^{\pm 0.3}$.
  \item The simulated $t$ dependence is varied by changing the slope 
parameters by $\pm 4 \ {\rm GeV^{-2}}$  for EL, $\pm 2 \ {\rm GeV^{-2}}$ for 
GD and PD and $\pm 1 \ {\rm GeV^{-2}}$ for DD processes.
  \item A $50\%$ variation is applied in the relative strengths of 
$\omega$ and $\phi$ production in the simulations.
  \item Uncertainties in the correction for migration between the two $\my$ 
intervals are estimated by varying the $Y$ fragmentation scheme assumed in
the simulations. An error is formed from the spread in corrections 
obtained using PHOJET, PYTHIA and DIFFVM in isotropic
and longitudinal phase space modes.
  \item The full
difference between results obtained when correcting the data with the PHOJET 
and PYTHIA simulations is taken as the uncertainty in the modelling of 
the final state.
  \item The errors on the luminosity measurements result in overall 
normalisation uncertainties of 1.7\% for the nominal vertex samples and 
5.4\% for the shifted vertex samples.  
  \item The uncertainty in the acceptance of the eTag results in overall
normalisation uncertainties of 3.5\% averaged over the range 
$164 < \Wgp < 212 \ {\rm GeV}$ and 6.0\% for
$212 < \Wgp < 251 \ {\rm GeV}$.
\end{itemize}
The systematic errors quoted for each data point 
are formed by adding in quadrature
the variations in the measured cross section arising from each uncertainty.

%% ========================== Results ==========================
\section{Results}
\label{results}
\subsection{The differential cross sections 
{\boldmath ${\rm d} \sigma / {\rm d} \mx^2$}}
\label{pointsec}
\begin{figure}[htb] \unitlength 1mm
 \begin{center}
  \begin{picture}(110,96)
   \put(0,-9){\epsfig{file=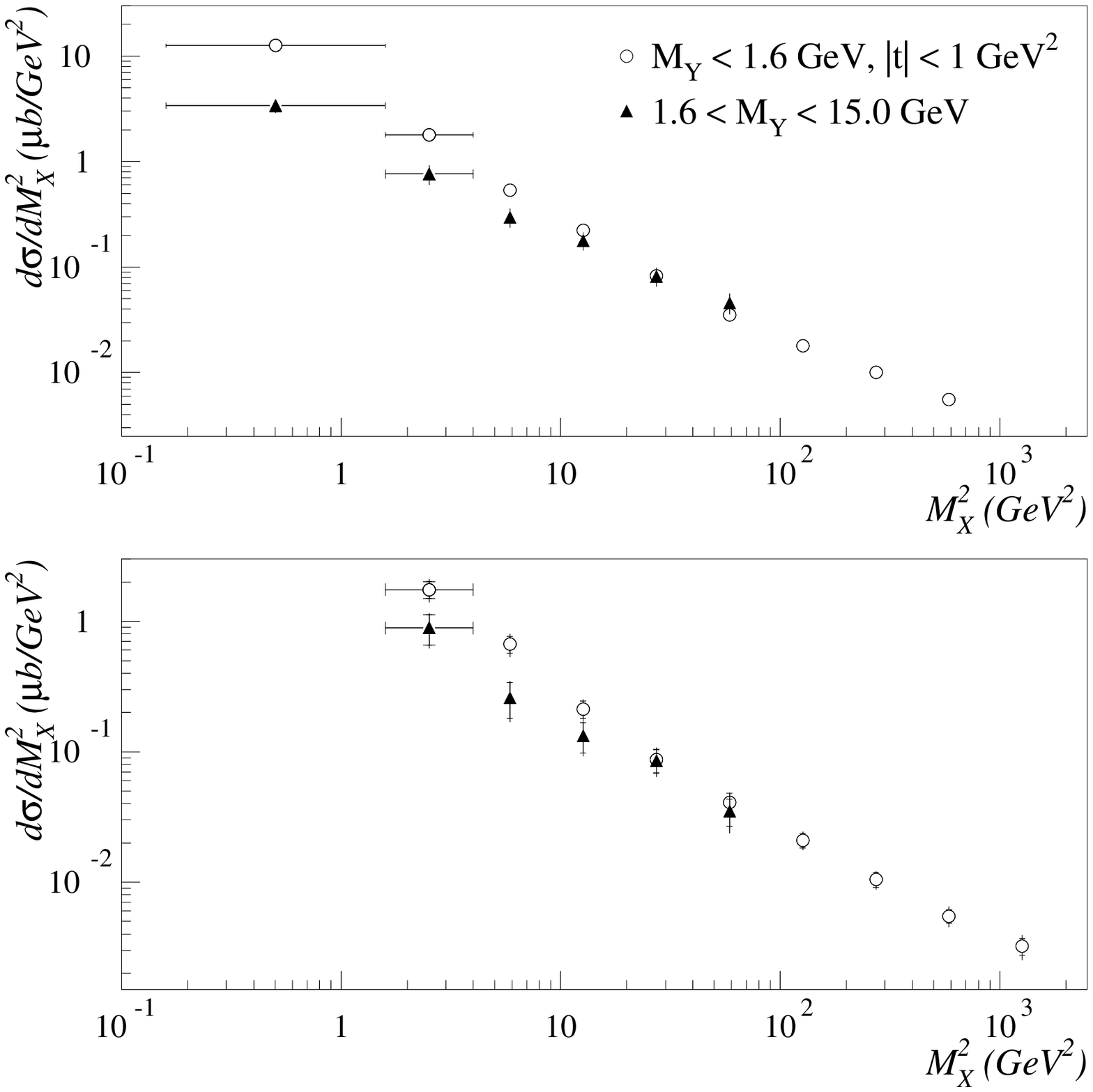,width=0.75\textwidth}}
   \begin{footnotesize}
     \put(20,8){(b) H1 Data $\av{\Wgp} = 231 \ {\rm GeV}$}
     \put(20,62){(a) H1 Data $\av{\Wgp} = 187 \ {\rm GeV}$}
   \end{footnotesize}
  \end{picture}
 \end{center}
 \vspace{-0.2cm}
  \scaption  {The measured differential cross section
${\rm d} \sigma / {\rm d} \mx^2$ for the process $\gamma p \rightarrow XY$ in
the regions $\my < 1.6 \ {\rm GeV}$, $\left| t \right| < 1.0 \ {\rm GeV^2}$
and $1.6 < \my < 15.0 \ {\rm GeV}$. (a) $\av{\Wgp} = 187 \ {\rm GeV}$, 
(b) $\av{\Wgp} = 231 \ {\rm GeV}$. The inner error bars show statistical
uncertainties only. The outer ones show statistical and systematic errors
added in quadrature.}
  \label{logplot}
\end{figure}

The measured differential cross sections are presented in
figure~\ref{logplot}. Figure~\ref{logplot}a shows 
${\rm d} \sigma_{\gamma p \rightarrow XY} / {\rm d}\mx^2$ 
at $\av{\Wgp} = 187 \ {\rm GeV}$, measured in 9 bins of $\mx^2$ in the range 
\linebreak 
$0.160 < \mx^2 < 862 \ {\rm GeV^2}$  with $\my < 1.6\ {\rm GeV}$ and 
$\left| t \right| < 1.0\ {\rm GeV^2}$. Superimposed is the cross section
integrated over the region $1.6 < \my < 15.0 \ {\rm GeV}$ and all $|t|$, 
in 6 $\mx^2$ intervals in the 
range $0.160 < \mx^2 < 86.2 \ {\rm GeV^2}$. 
The first interval in $\mx^2$ is chosen to cover the region of
$\rho^0$, $\omega$ and $\phi$ production. The second is constructed to 
contain the $\rho^{\prime}$ resonances~\cite{rhoprime}. For these two
measurements an error bar, reflecting the width of the $\mx^2$
interval, is shown to account for 
uncertainties in the details of the resonance structure.\footnote{In the 
remaining $\mx^2$ intervals the largest deviation from a smooth dependence 
on $\mx^2$ is likely to arise from
$J/\psi$ production, which contribute at the level of $5 \%$ in the 
fourth $\mx^2$ interval~\cite{h1:jpsi}.}
For all but the 
first two $\mx^2$ intervals measurements are made in three bins per decade of 
$\mx^2$. The results of the measurement at 
$\av{\Wgp} = 231 \ {\rm GeV}$ are shown in figure~\ref{logplot}b.
All cross section measurements are summarised in table~\ref{gdpoints}.

\begin{table}[h]
\begin{center}
{\bf (a)}
\begin{footnotesize}
\begin{tabular}{|c|l||cccl||cccl|} \cline{3-10}
\multicolumn{2}{c|}{} & \multicolumn{4}{c||}{$\av{\Wgp} = 187 \ {\rm GeV}$} & \multicolumn{4}{c|}{$\av{\Wgp} = 231 \ {\rm GeV}$} \\ \hline
$\mx^2$ & Bin Limits & $\frac{{\rm d} \sigma}{{\rm d} \mx^2}$  & Stat. & Syst. & & $\frac{{\rm d} \sigma}{{\rm d} \mx^2}$  & Stat. & Syst. & \\ 
(${\rm GeV^2}$) & \hspace{0.15cm} (${\rm GeV^2}$) & \multicolumn{3}{c}{(${\rm \mu b \ GeV^{-2}}$)} & & \multicolumn{3}{c}{(${\rm \mu b \ GeV^{-2}}$)} & \\ \hline
0.504 & $0.16 - 1.58$ &  ($1.25$ & $\pm 0.05$ & $\pm 0.11$) & \hspace*{-0.35cm} $\times 10^{1}$  & \multicolumn{4}{c|}{---}  \\ \hline 
2.52  & $1.58 - 4.00$ &  ($1.80$ & $\pm 0.11$ & $\pm 0.22$) & \hspace*{-0.35cm} $\times 10^{0}$  & ($1.75$ & $\pm 0.26$ & $\pm 0.24$) & \hspace*{-0.35cm} $\times 10^{0}$ \\ \hline
5.87  & $4.00 - 8.62$ &  ($5.37$ & $\pm 0.34$ & $\pm 0.48$) & \hspace*{-0.35cm} $\times 10^{-1}$ & ($6.67$ & $\pm 1.00$ & $\pm 0.92$) & \hspace*{-0.35cm} $\times 10^{-1}$ \\ \hline 
12.7  & $8.62 - 18.6$ &  ($2.24$ & $\pm 0.15$ & $\pm 0.22$) & \hspace*{-0.35cm} $\times 10^{-1}$ & ($2.13$ & $\pm 0.32$ & $\pm 0.24$) & \hspace*{-0.35cm} $\times 10^{-1}$ \\ \hline 
27.3  & $18.6 - 40.0$ &  ($8.25$ & $\pm 0.74$ & $\pm 0.52$) & \hspace*{-0.35cm} $\times 10^{-2}$ & ($8.68$ & $\pm 1.76$ & $\pm 1.12$) & \hspace*{-0.35cm} $\times 10^{-2}$ \\ \hline 
58.7  & $40.0 - 86.2$ &  ($3.53$ & $\pm 0.34$ & $\pm 0.25$) & \hspace*{-0.35cm} $\times 10^{-2}$ & ($4.07$ & $\pm 0.72$ & $\pm 0.36$) & \hspace*{-0.35cm} $\times 10^{-2}$ \\ \hline 
127   & $86.2 - 186$ &  ($1.79$ & $\pm 0.15$ & $\pm 0.19$) & \hspace*{-0.35cm} $\times 10^{-2}$ & ($2.10$ & $\pm 0.28$ & $\pm 0.21$) & \hspace*{-0.35cm} $\times 10^{-2}$ \\ \hline 
273   & $186 - 400$  &  ($1.01$ & $\pm 0.76$ & $\pm 0.91$) & \hspace*{-0.35cm} $\times 10^{-2}$ & ($1.05$ & $\pm 0.14$ & $\pm 0.10$) & \hspace*{-0.35cm} $\times 10^{-2}$ \\ \hline 
587   & $400 - 862$  &  ($5.56$ & $\pm 0.40$ & $\pm 0.55$) & \hspace*{-0.35cm} $\times 10^{-3}$ & ($5.50$ & $\pm 0.65$ & $\pm 0.73$) & \hspace*{-0.35cm} $\times 10^{-3}$ \\ \hline 
1270  & $862 - 1860$  & \multicolumn{4}{c||}{---}                            & ($3.22$ & $\pm 0.48$ & $\pm 0.52$) & \hspace*{-0.35cm} $\times 10^{-3}$ \\ \hline 
\end{tabular}
\end{footnotesize}
\end{center}
\begin{center}
{\bf (b)} 
\begin{footnotesize}
\begin{tabular}{|c|l||cccl||cccl|} \cline{3-10}
\multicolumn{2}{c|}{} & \multicolumn{4}{c||}{$\av{\Wgp} = 187 \ {\rm GeV}$} & \multicolumn{4}{c|}{$\av{\Wgp} = 231 \ {\rm GeV}$} \\ \hline
$\mx^2$ & Bin Limits & $\frac{{\rm d} \sigma}{{\rm d} \mx^2}$  & Stat. & Syst. & & $\frac{{\rm d} \sigma}{{\rm d} \mx^2}$  & Stat. & Syst. & \\ 
(${\rm GeV^2}$) & \hspace*{0.15cm} (${\rm GeV^2}$) & \multicolumn{3}{c}{(${\rm \mu b \ GeV^{-2}}$)} & & \multicolumn{3}{c}{(${\rm \mu b \ GeV^{-2}}$)} & \\ \hline
0.504 & $0.16 - 1.58$ &  ($3.40$ & $\pm 0.25$ & $\pm 0.40$) & \hspace*{-0.35cm} $\times 10^{0}$  & \multicolumn{4}{c|}{---}  \\ \hline 
2.52  & $1.58 - 4.00$ &  ($7.60$ & $\pm 0.80$ & $\pm 1.38$) & \hspace*{-0.35cm} $\times 10^{-1}$ & ($8.91$ & $\pm 2.34$ & $\pm 1.34$) & \hspace*{-0.35cm} $\times 10^{-1}$ \\ \hline
5.87  & $4.00 - 8.62$ &  ($2.98$ & $\pm 0.34$ & $\pm 0.49$) & \hspace*{-0.35cm} $\times 10^{-1}$ & ($2.59$ & $\pm 0.79$ & $\pm 0.46$) & \hspace*{-0.35cm} $\times 10^{-1}$ \\ \hline 
12.7  & $8.62 - 18.6$ &  ($1.79$ & $\pm 0.18$ & $\pm 0.30$) & \hspace*{-0.35cm} $\times 10^{-1}$ & ($1.33$ & $\pm 0.35$ & $\pm 0.20$) & \hspace*{-0.35cm} $\times 10^{-1}$ \\ \hline 
27.3  & $18.6 - 40.0$ &  ($8.19$ & $\pm 0.82$ & $\pm 1.44$) & \hspace*{-0.35cm} $\times 10^{-2}$ & ($8.57$ & $\pm 1.79$ & $\pm 1.22$) & \hspace*{-0.35cm} $\times 10^{-2}$ \\ \hline 
58.7  & $40.0 - 86.2$ &  ($4.56$ & $\pm 0.51$ & $\pm 0.85$) & \hspace*{-0.35cm} $\times 10^{-2}$ & ($3.51$ & $\pm 0.84$ & $\pm 0.79$) & \hspace*{-0.35cm} $\times 10^{-2}$ \\ \hline 
\end{tabular}
\end{footnotesize}
\end{center}
\vspace{-0.3cm}
\scaption{The measured differential cross section 
${\rm d} \sigma / {\rm d} \mx^2$ for 
the ranges (a) $\my < 1.6 \ {\rm GeV}$ and 
$\left| t \right| < 1.0 \ {\rm GeV^2}$ and 
(b) $1.6 < \my < 15.0 \ {\rm GeV}$ and all $t$.
All measurements are quoted at the bin centres in $\log
\mx^2$. Overall scale uncertainties of $5.2 \%$ at 
$\av{\Wgp} = 187 \ {\rm GeV}$ and $6.9 \%$ at $\av{\Wgp} = 231 \ {\rm GeV}$
are not included in the systematic errors.}
\label{gdpoints}
\end{table}

At sufficiently large $\mx^2$ the measurements for $\my < 1.6 \ {\rm GeV}$ 
in both $\Wgp$ ranges display an
approximate ${\rm d} \sigma / {\rm d} \mx^2 \sim 1 / \mx^2$ dependence,
though closer scrutiny reveals a more complex structure
which is investigated in section~\ref{fitsec}. 
The $\mx^2$ distribution for \linebreak $1.6 < \my  < 15.0 \ {\rm GeV}$ 
falls slightly 
less steeply with increasing $\mx^2$ than that for $\my < 1.6 \ {\rm GeV}$.
The cross section is
enhanced considerably at the lowest values of $\mx^2$ and $\my^2$
where elastic processes dominate~\cite{h1:rho}. 

Previous data exist on the differential cross section for $\gamma p
\rightarrow Xp$~\cite{E612:gp}.  Since the region $\my < 1.6 \ {\rm GeV}$ is
expected to be heavily dominated by this process, the low $\my$ measurement
is compared to fixed target data in section~\ref{fitsec} and combined fits
are performed using the Regge pole model described in
section~\ref{formalism}.  The degree to which diffractive and non-diffractive
exchanges contribute to the single photon dissociation process is
investigated.  The questions are addressed as to whether the intercept of the
pomeron in such a model is consistent with that previously deduced from  soft
hadron-hadron and photoproduction
interactions and whether any anomalous behaviour
is necessary to explain the $\Wgp$ dependence
of the diffractive contribution.  The behaviour of the cross section for $1.6
< \my < 15.0 \ {\rm GeV}$ is discussed in section~\ref{ddsec}.

\subsection{Triple-Regge model of photon dissociation}
\label{formalism}

\begin{figure}[h] \unitlength 1mm
 \begin{center}
   \begin{picture}(160,45)
     \put(3,10){\epsfig{file=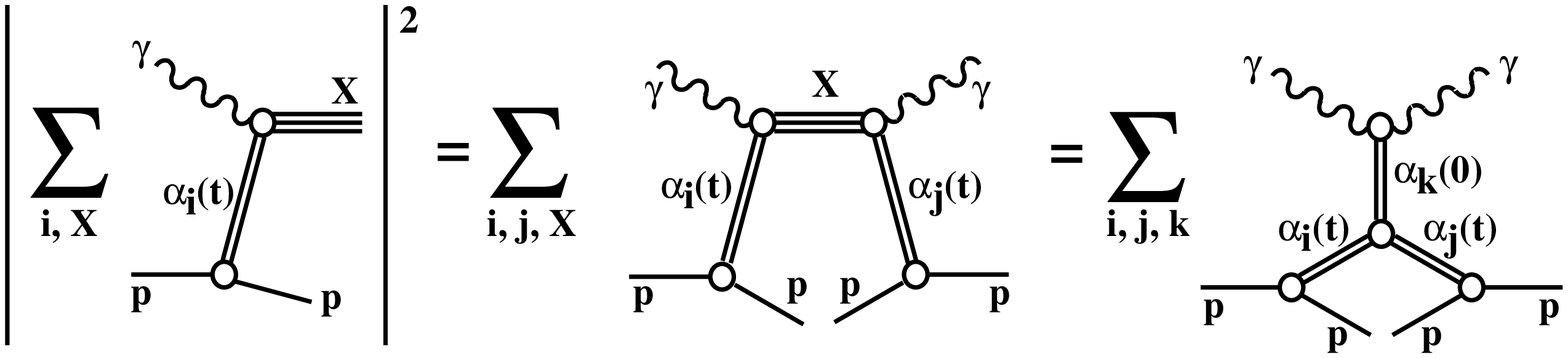,width=0.95\textwidth}}
     \put(20,3){\bf{(a)}}
     \put(79,3){\bf{(b)}}
     \put(134,3){\bf{(c)}}
   \end{picture}
 \end{center}
 \vspace{-0.6cm}
  \scaption  {Illustration of the Mueller-Regge approach to the inclusive
photon dissociation cross section.}
  \label{tripreg}
\end{figure}

An appropriate framework in which to model dissociation processes is offered
by `triple-Regge' phenomenology. For $\Wgp^2 \gg \mx^2$ the process
$\gamma p \rightarrow Xp$ may be treated with a Regge expansion. The
amplitude at fixed $\mx$ is then a sum of amplitudes for the exchange of
reggeons $i$ that produce all possible final states $X$, as illustrated in
figure~\ref{tripreg}a. The corresponding cross section contains products of
flux factors for reggeons $i$ and $j$ and amplitudes $T_{\gamma \alpha_i(t)
  \rightarrow X} \ T^{\star}_{\gamma \alpha_j(t) \rightarrow X}$
(figure~\ref{tripreg}b).  The generalised optical theorem~\cite{mueller:opt} 
relates the sum
over $X$ of these matrix elements to the forward amplitude for the process
$\gamma \alpha_i (t) \rightarrow \gamma \alpha_j (t)$ 
\linebreak at an effective centre
of mass energy $\mx$. When $\mx^2$ is large by comparison with the hadronic
mass scale $s_{_0}$, a Regge expansion is
also appropriate for the photon-reggeon scattering amplitude, such that the
dissociation cross section may be decomposed into triple-Regge terms as shown
in figure~\ref{tripreg}c.
The cross section may then be expressed as a sum of
contributions with reggeons $i$, $j$ and
$k$~\cite{kaidalov:review,alberi:goggi,goulianos:review,zotov:tsarev}
\begin{eqnarray}
  \frac{ {\rm d} \sigma}{{\rm d}t \, {\rm d}\mx^2} = \frac{s_{_0}}{\Wgp^4} 
  \sum_{i,j,k} G_{ijk}(t) \
  \left(\frac{\Wgp^2}{\mx^2}\right)^{\alpha_i(t) + \alpha_j(t)} 
  \left(\frac{\mx^2}{s_{_0}} \right)^{\alpha_k(0)} \ 
  \cos \left[ \phi_i(t) - \phi_j(t) \right] \ .
  \label{trbasic}
\end{eqnarray}

The functions $G_{ijk}(t)$ and $\alpha_i(t)$ are not predicted by the model
and must be determined from experimental measurements.
The trajectories $\alpha_i (t)$
are assumed to take the linear form $\alpha_i (t) = \alpha_i (0) + 
\alpha_i^{\prime} t$. The phase $\phi_i(t)$ of reggeon $i$ is determined by the
signature factor, $\eta_i(t) = \zeta + e^{- i \pi \alpha_i(t)}$, where $\zeta
= \pm 1$ is the signature of the exchange. The signature factors
are written as $\eta_i(t) = \eta_i^0(t) \ e^{i \phi_i(t)}$ with the moduli
$\eta_i^0 (t)$ absorbed into the $\beta$ parameters introduced in
equation (\ref{trcoup}).
For photoproduction reggeons $i$ and $j$ must have
the same signature such that $\phi_i(t) - \phi_j(t) = \frac{\pi}{2}
\left[ \alpha_j(t) - \alpha_i(t) \right]$.  As is customary 
the scale $s_{_0}$ is set to $1\ {\rm GeV^2}$.  
The functions $G_{ijk}(t)$ may be factored into 
products of couplings each of the form\footnote{The couplings $G_{ijk}(t)$ 
are expected to differ from
those extracted from $pp$ data by a factor 
$ \beta_{\gamma k}(0)/ \beta_{p k}(0)$.}
\begin{eqnarray}
  G_{ijk}(t) = \frac{1}{16 \pi} \ \beta_{p i}(t) \ \beta_{p j}(t) \ 
    \beta_{\gamma k}(0) \ g_{ijk}(t) \ ,
  \label{trcoup}
\end{eqnarray}
where the $\beta$ terms describe the couplings of the reggeons to external
particles and \linebreak
$g_{ijk}(t)$ is the appropriate three-reggeon coupling. 
The $t$ dependence of the reggeon-proton and three-reggeon couplings are
parameterised here as
$\beta_{pi}(t) = \beta_{pi}(0) e^{b_{p i} t}$ and
$g_{ijk}(t) = g_{ijk}(0) e^{b_{ijk} t}$. 

The pomeron, with trajectory $\alphapom (t)$, is unique in Regge theory in 
that its intercept
is significantly larger than those of all other reggeons. In the
limit in which both $\Wgp^2 / \mx^2$ and $\mx^2 \rightarrow \infty$ only the
$ijk = \pom \pom \pom$ term survives and
equation (\ref{trbasic}) reduces to
\begin{eqnarray}
  \frac{ {\rm d} \sigma}{{\rm d}t \, {\rm d}\mx^2} =
  \frac{G_{\pom \pom \pom}(0)}{s_{_0}^{\alphapom(0) - 1}} \
  \left( \Wgp^2 \right)^{2 \alphapom(0) - 2} \
  \left( \frac{1}{\mx^2} \right)^{\alphapom(0)} \
  e^{B (\Wgp^2, \mx^2) \: t} \ ,
  \label{pomonly}
\end{eqnarray}
where $B (\Wgp^2, \mx^2) = 2 b_{p \pom} + b_{\pom \pom \pom} + 2
\alphapom^{\prime} \ln \Wgp^2 / \mx^2$.  After the pomeron the next-leading
reggeons have approximately degenerate trajectories and carry the quantum
numbers of the $\rho$, $\omega$, $a_2$ and $f_2$ mesons.\footnote{The
reggeons under consideration are hereafter referred to as $\pom$, $\rho$,
$\omega$, $a$ and $f$. Their
isospin, signature and C- and G-parities are
$\pom (0+\!++)$, $\rho (1-\!-+)$, $\omega (0-\!--)$, $a (1+\!+-)$ and 
$f (0+\!++)$.} In this analysis the symbol $\reg$ is used to describe
combinations of these four reggeons and a single effective trajectory
$\alphareg(t)$ is assumed. 

With the two trajectories, $\alphapom(t)$ and
$\alphareg(t)$, equation (\ref{trbasic})
leads to a total of six terms with distinct $\Wgp^2$ and $\mx^2$ dependences.
Table~\ref{trterms} lists each of these terms.  Diffractive contributions are
approximately independent of $\Wgp^2$ and correspond to the case in
figure~\ref{tripreg}c where both reggeons $i$ and $j$ are the pomeron. In
addition to the triple-pomeron diagram, a further diffractive term arises from
the $ijk = \pom \pom \reg$ 
diagram. The reggeon $k$ must have the quantum numbers
of the $f$ in order to satisfy the requirements of
conservation of C-parity at the photon
vertex and C- and G-parity at the three-reggeon vertex.

The two non-diffractive terms, $\reg \reg \pom$ and $\reg \reg \reg$,
may involve the exchange of any of $\rho$, $\omega$, $a$ or $f$ and are
suppressed by factors like $1/\Wgp^2$ at fixed $\mx^2$. The production of 
states $Y$ that have different isospin or charge from the proton
($Y = n$, $N^{\star \, 0}$, $\Delta^{++}$, $\Delta^{+}$, $\Delta^0$, $\ldots$)
can only occur via isovector exchanges
($i = j = \rho$ or $a$).
Such processes are described, up to a normalisation, in
the same way as processes that are elastic at the proton vertex, 
through those specific terms in table~\ref{trterms} marked with a 
star.

Since the $\pom$ and $f$ reggeons exchange identical quantum numbers 
they may 
inter-\linebreak fere~\cite{field:fox,dl:fdom,inami:roberts,
kazarinov,hidaka}.\footnote{Only
interference between the $\pom$ and $f$ reggeons is considered in this
analysis. Since the
photon is not an eigenstate of G-parity, interference between
reggeons of opposite G-parity is in principle also possible. However, the
dominance of two-pion final states at small $\mx$ suggests that the photon
is dominated by ${\rm G} = +1$ and that the interference terms 
$\pom {\em a a}$, $\rho \omega a$ and $f a a$ are likely to be 
small \cite{WATSON}.}
Such a contribution is expressed in terms of the `non-diagonal' amplitudes, 
$\{\pom \reg \} \pom$ and $\{ \pom \reg \} \reg$ where $\{ \pom \reg \} 
{\em k} \equiv \pom \reg {\em k} + \reg \pom {\em k}$.  
The $f$ reggeon has previously been 
found to couple strongly to the proton~\cite{dl:fdom} and since the 
interference terms have weaker de\-pend\-ences on $\Wgp^2$ 
than the non-diffractive contributions ($\reg \reg \pom$ and $\reg \reg \reg$),
they
may be of particular significance in the high energy regime of the present
measurement. 
Restrictions are imposed on the magnitudes of the interference
relative to the non-interference terms through the
inequality \cite{roberts:roy}
\begin{eqnarray}
 \left\{
 \sum_k
 G_{_{\{ \pom \reg \} k}}(t) \left(\mx^2 \right)^{\alpha_k(0)} \right\}^2 
 \leq 4 \ \sum_k
 G_{_{\pom \pom k}}(t) \left(\mx^2 \right)^{\alpha_k(0)} \
 \sum_k
 G_{_{\reg \reg k}}(t) \left(\mx^2 \right)^{\alpha_k(0)} \ .
 \label{schwarz}
\end{eqnarray}
\begin{table}[h]
\begin{center}
\begin{footnotesize}
\begin{tabular}{|c|c||c|c|} \hline
Generic & Specific & Approx. $\Wgp$ & Approx. $\mx$  \\
term $ijk$ & terms $ijk$ & dependence & dependence \\ \hline 
$\pom \pom \pom$       & $\pom \pom \pom$ & $\Wgp^0$ & $\mx^{-2}$ \\ \hline 
$\pom \pom \reg$       & $\pom \pom {\em f}$    & $\Wgp^0$ & $\mx^{-3}$ \\ \hline 
$\reg \reg \pom$       & $f f \pom$ \ $a a \pom^{\star}$ \ $\rho \rho \pom^{\star}$ \ $\omega \omega \pom$ & $\Wgp^{-2}$ & $\mx^0$ \\ \hline 
$\reg \reg \reg$       & $f f f$ \ $a a f^{\star}$ \ $\rho \rho f^{\star}$ \ $\omega \omega f$ \ $\{f a \} a$ \ $\{ \rho \omega \} a$ & $\Wgp^{-2}$ & $\mx^{-1}$ \\ \hline 
$\{ \pom \reg \} \pom$ & $\{ \pom {\em f} \} \pom$ & $\Wgp^{-1}$ & $\mx^{-1}$ \\ \hline 
$\{ \pom \reg \} \reg$ & $\{ \pom {\em f} \} {\em f}$ \ $\{ \pom {\em a} \} {\em a}$ & $\Wgp^{-1}$ & $\mx^{-2}$ \\ \hline 
\end{tabular}
\end{footnotesize}
\end{center}
\vspace{-0.3cm}
\scaption{The $\Wgp$ and $\mx$ dependence of each of the six 
triple-Regge amplitudes considered in the fits under the approximations
$\alphapom(t) \simeq 1$ and $\alphareg(t) \simeq 0.5$. The second column lists
each of the particular terms that are allowed by quantum number conservation.
Those marked with a star can result in states $Y$ that have different isospin
or charge from the proton.}
\label{trterms}
\end{table}

The intercept of the pomeron is a free parameter in the fits presented here. 
A free fit for all other parameters is not practical with the limited
photoproduction data available. Instead, information is taken from data on
total and elastic cross sections. The intercept 
of the subleading trajectory is taken
from the results of fits~\cite{dl:stot} to the centre of mass energy 
($\sqrt{s}$) dependence of total cross sections that use
expressions of the form
\begin{eqnarray}
  \sigma_{\rm tot}^{ab} = 
  A^{ab} s^{\alpha_{_{\pom}}(0) - 1} +
  B^{ab} s^{\alpha_{_{\reg}}(0) - 1} \ .
  \label{sigmatot}
\end{eqnarray}
The trajectory gradient $\alphapom^{\prime}$ is obtained from the shrinkage
of the forward peak in high energy elastic $\bar{p} p$
scattering~\cite{cdf:elas}. The slopes for all
non-diffractive trajectories are assumed to be the same and the value of
$\alphareg^{\prime}$ is taken from the result obtained for
$\alpha_\rho^{\prime}$ in analyses~\cite{apel:chexch}
of the reaction $\pi^- p \rightarrow \pi^0 n$. The slope parameter
$b_{p \pom}$ is determined from elastic scattering
measurements~\cite{cdf:elas} in the regime of diffractive dominance using
\begin{eqnarray}
  \frac{{\rm d} \sigma_{_{\rm EL}}^{pp}}{{\rm d} t} = 
  \left( \frac{{\rm d} \sigma_{_{\rm EL}}^{pp}}{{\rm d} t} \right)_{t = 0} 
  \ e^{B(s) \: t} \ ,
  \label{sigmael}
\end{eqnarray}
where $B(s) = 4 b_{p \pom} + 2 \alpha_{_{\pom}}^{\prime} \ln s / s_0$. The
parameter $b_{p \reg}$ is less well determined and is allowed to vary within
a range that
covers most of the results obtained in lower energy fits.  Previous
meas\-ure\-ments~\cite{cool:dude} have found that the $t$-dependence of the
triple-pomeron coupling is weak compared to those of the external
vertices, though less is known about the remaining three-reggeon couplings.
It is assumed here that the dominant $t$ dependence of $G_{ijk}(t)$ arises
from the $\beta$ terms and uncertainties in the $t$ dependence of the
three-reggeon coup\-lings are accounted for by setting $b_{ijk} = 0 \pm 1 \ 
{\rm GeV^{-2}}$ for all $ijk$.  Table~\ref{params} specifies the values
assumed for the parameters that are kept fixed in the fits and the assigned
uncertainties.

\begin{table}[h]
\begin{center}
\begin{footnotesize}
\begin{tabular}{|c||lll||c|} \hline
Quantity & \multicolumn{3}{c}{Value} & Source \\ \hline 
$\alpha_{_{\reg}}(0)$       & $0.55$ & $\pm 0.10$ &               & \cite{dl:stot} \\ \hline
$\alpha_{_{\pom}}^{\prime}$ & $0.26$ & $\pm 0.02$ & \hspace{-0.3cm} ${\rm GeV^{-2}}$ & \cite{cdf:elas} \\ \hline 
$\alpha_{_{\reg}}^{\prime}$ & $0.90$ & $\pm 0.10$ & \hspace{-0.3cm} ${\rm GeV^{-2}}$ & \cite{apel:chexch} \\ \hline 
$b_{p \pom}$                & $2.3$  & $\pm 0.3$  & \hspace{-0.3cm} ${\rm GeV^{-2}}$ & \cite{cdf:elas} \\ \hline 
$b_{p \reg}$                & $1.0$  & $\pm 1.0$  & \hspace{-0.3cm} ${\rm GeV^{-2}}$ & \cite{kaidalov:review} \\ \hline 
$b_{ijk}$                   & $0.0$  & $\pm 1.0$  & \hspace{-0.3cm} ${\rm GeV^{-2}}$ & \\ \hline 
\end{tabular}
\end{footnotesize}
\end{center}
\vspace{-0.3cm}
\scaption{Values assumed for the fixed parameters in the triple-Regge fits and
the sources from which they are taken. Each quantity is defined in the text.
Contributions to the model dependence errors are formed by repeating the fits
after separately varying each parameter by the quoted uncertainties.}
\label{params}
\end{table}

\subsection{Regge analysis of the proton-elastic cross section}
\label{fitsec}

Large ranges in both $\Wgp^2$ and $\mx^2$ are required in order to investigate
the importance of all six terms in table~\ref{trterms}.  To extend the range
in $\Wgp^2$ simultaneous fits are performed to H1 and to fixed target
data~\cite{E612:gp}, where the quantity $\frac{{\rm d} \sigma (\gamma p
  \rightarrow Xp)}{{\rm d} t \, {\rm d} \mx^2}$ was presented at $t = -0.05 \ 
{\rm GeV^2}$ in the ranges $11.9 < \Wgp < 13.7 \ {\rm GeV}$ and $13.7 < \Wgp
< 16.7 \ {\rm GeV}$.  For the data from~\cite{E612:gp} equation
(\ref{trbasic}) is used at $t = -0.05 \ {\rm GeV^2}$. For the H1
data equation (\ref{trbasic}) is integrated over the measured range, $\left|
t_{\rm min} \right| < \left| t \right| < 1\ {\rm GeV^2}$, where $\left|
t_{\rm min} \right|$ is the minimum kinematically accessible value of $\left
| t \right |$~\cite{data:book}.  The parameters listed in table~\ref{params}
are fixed and $\alphapom (0)$ and the four couplings, 
$G_{_{\pom \pom \pom}}(0)$, $G_{_{\pom \pom \reg}}(0)$,
$G_{_{\reg \reg \pom}}(0)$ and $G_{_{\reg \reg \reg}}(0)$,
are free fit parameters. In order to avoid the complications
of the resonance region only data for which $\mx^2 > 4 \ {\rm GeV^2}$ are
included in the fits.

The additional contribution in the H1 data from events in which
the proton dissociates with $\my < 1.6 \ {\rm GeV}$ is accounted for
by applying a factor of $1.10 \pm 0.06$ to equation (\ref{trbasic})
in the triple-Regge analysis for
the H1 measurements only\footnote{No correction is necessary for the fixed
target measurements, which were performed by tagging leading final state
protons.  There may be a contribution from protons arising from decays of
$\Delta$ and other resonances, though this effect is expected to be
negligible~\cite{kazarinov}.}. This correction is determined by taking an
average of the values expected in the PHOJET and PYTHIA models after weights
have been applied which make the differential cross sections in the models
agree with those measured (see section~\ref{mcmods}). 
The uncertainty in this correction is taken as the
difference in the values obtained from the two models.
The correction does not account for additional states $Y$ that may be
produced by isovector exchange, but cannot be produced by diffractive or
other isoscalar
exchange. Any contribution from these states would increase the
couplings $G_{_{\reg \reg \pom}} (0)$ and $G_{_{\reg \reg \reg}} (0)$ for the
H1 measurement relative to the fixed target data (see table~\ref{trterms}).

No good solutions are found when diffractive terms only are considered. In a
fit in which the pomeron alone contributes, equivalent to equation 
(\ref{pomonly}) with $\alphapom(0)$ and $G_{_{\pom \pom \pom}}$
as free parameters, the $\chi^2 / {\rm ndf}$ made from statistical errors is
$89.9 / 33$. When both
terms $\pom \pom \pom$ and $\pom \pom \reg$ are included, the result is
$\chi^2 / {\rm ndf} = 79.3/32$. We conclude that significant non-diffractive
contributions are needed to explain the measured cross sections.

\begin{table}[h]
%\begin{center}
\begin{footnotesize}
\hspace{3cm} {\bf (a)} $\chi^2 / {\rm ndf} = 28.8 / 30$ \\
\begin{tabular}{|c|lllll|r|r|r|r|} \cline{7-10}
\multicolumn{6}{c|}{} & \multicolumn{4}{c|}{Correlation coefficients} \\ \hline
Parameter & Value & Stat. & Syst. & Model & & $G_{_{\pom \pom \pom}}$ & $G_{_{\pom \pom \reg}}$ & $G_{_{\reg \reg \pom}}$ & $G_{_{\reg \reg \reg}}$ \\ \hline
$\alphapom(0)$             & 1.031  & $\pm 0.014$ & $\pm 0.013$ & $\pm 0.009$ &                       & $-0.93$ & $-0.86$ & $-0.74$ & $0.82$   \\ \hline
$G_{_{\pom \pom \pom}}(0)$ & $8.19$ & $\pm 1.60$  & $\pm 1.34$  & $\pm 2.22$  & \hspace*{-0.6cm} ${\rm \mu b \, GeV^{-2}}$ &        & $0.63$  & $0.50$  & $-0.60$  \\ \hline
$G_{_{\pom \pom \reg}}(0)$ & $14.0$ & $\pm 5.2$  & $\pm 8.1$    & $\pm 2.2$  & \hspace*{-0.6cm} ${\rm \mu b \, GeV^{-2}}$ &        &         & $0.89$  & $-0.94$  \\ \hline
$G_{_{\reg \reg \pom}}(0)$ & $238$  & $\pm 73$    & $\pm 98$    & $\pm 101$    & \hspace*{-0.6cm} ${\rm \mu b \, GeV^{-2}}$ &       &         &       & $-0.98$  \\ \hline
$G_{_{\reg \reg \reg}}(0)$ & $-506$ & $\pm 293$   & $\pm 325$   & $\pm 328$    & \hspace*{-0.6cm} ${\rm \mu b \, GeV^{-2}}$ &       &         &       &         \\ \hline
\end{tabular} 
\vskip 0.3cm
\hspace{3cm} {\bf (b)} $\chi^2 / {\rm ndf} = 19.9 / 30$ \\
\begin{tabular}{|c|lllll|r|r|r|r|} \cline{7-10}
\multicolumn{6}{c|}{} & \multicolumn{4}{c|}{Correlation coefficients} \\ \hline
Parameter & Value & Stat. & Syst. & Model & & $G_{_{\pom \pom \pom}}$ & $G_{_{\pom \pom \reg}}$ & $G_{_{\reg \reg \pom}}$ & $G_{_{\reg \reg \reg}}$ \\ \hline
$\alphapom(0)$             & 1.101  & $\pm 0.010$ & $\pm 0.022$ & $\pm 0.022$ &                       & $-0.88$ & $-0.46$ & $0.23$  & $0.08$   \\ \hline
$G_{_{\pom \pom \pom}}(0)$ & $2.05$ & $\pm 0.44$  & $\pm 1.26$  & $\pm 1.21$  & \hspace*{-0.6cm} ${\rm \mu b \, GeV^{-2}}$ &        & $0.06$  & $-0.59$ & $0.32$   \\ \hline
$G_{_{\pom \pom \reg}}(0)$ & $4.19$ & $\pm 0.76$  & $\pm 1.36$    & $\pm 1.70$  & \hspace*{-0.6cm} ${\rm \mu b \, GeV^{-2}}$ &        &      & $0.42$  & $-0.64$  \\ \hline
$G_{_{\reg \reg \pom}}(0)$ & $115$  & $\pm 29$    & $\pm 65$    & $\pm 78$    & \hspace*{-0.6cm} ${\rm \mu b \, GeV^{-2}}$ &       &         &       & $-0.93$  \\ \hline
$G_{_{\reg \reg \reg}}(0)$ & $-405$ & $\pm 217$   & $\pm 382$   & $\pm 464$    & \hspace*{-0.6cm} ${\rm \mu b \, GeV^{-2}}$ &       &         &       &         \\ \hline
\end{tabular}
\end{footnotesize}
\vskip 0.3cm
\begin{footnotesize}
\hspace{3cm} {\bf (c)} $\chi^2 / {\rm ndf} = 18.4 / 29$ \\
\begin{tabular}{|c|lllll|r|r|r|r|r|} \cline{7-11}
\multicolumn{6}{c|}{} & \multicolumn{5}{c|}{Correlation coefficients} \\ \hline
Parameter & Value & Stat. & Syst. & Model & & $G_{_{\pom \pom \pom}}$ & $G_{_{\pom \pom \reg}}$ & $G_{_{\reg \reg \pom}}$ & $G_{_{\reg \reg \reg}}$ & 
\multicolumn{1}{c|}{${\cal R}$} \\ \hline
$\alphapom(0)$                       & 1.071  & $\pm 0.024$ & $\pm 0.021$ & $\pm 0.018$ &                       & $-0.98$ & $-0.93$ & $-0.85$ & $0.92$ & $0.80$   \\ \hline
$G_{_{\pom \pom \pom}}(0)$           & $3.76$ & $\pm 1.62$  & $\pm 2.25$  & $\pm 1.45$  & \hspace*{-0.6cm} ${\rm \mu b \, GeV^{-2}}$ &        & $0.83$  & $0.83$ & $-0.89$ & $-0.82$   \\ \hline
$G_{_{\pom \pom \reg}}(0)$           & $7.46$ & $\pm 3.73$  & $\pm 2.81$   & $\pm 2.67$   & \hspace*{-0.6cm} ${\rm \mu b \, GeV^{-2}}$     &        &        & $0.84$ & $-0.90$ & $-0.75$    \\ \hline
$G_{_{\reg \reg \pom}}(0)$           & $63.8$    & $\pm 70.7$    & $\pm 17.7$    & $\pm 39.7$ & \hspace*{-0.6cm} ${\rm \mu b \, GeV^{-2}}$ &        &        &       & $-0.99$ & $-0.98$   \\ \hline
$G_{_{\reg \reg \reg}}(0)$           & $264$    & $\pm 344$    & $\pm 27$    & $\pm 159$    & \hspace*{-0.6cm} ${\rm \mu b \, GeV^{-2}}$ &        &        &        &    & $0.95$   \\ \hline
${\cal R}$    & $4.56$  & $\pm 3.60$    & $\pm 1.74$    & $\pm 1.41$   &  &        &        &        &        &  \\ \hline
\end{tabular}
\end{footnotesize}
\vspace{-0.1cm}
\scaption{Results of the triple-Regge fits and statistical, systematic and 
model related errors. The $\chi^2 / {\rm ndf}$ values and correlation
coefficients reflecting statistical errors only are also given. 
(a) Fit without interference or isovector exchange contributions.
(b) Maximal constructive interference 
between the diffractive and the secondary exchange. (c) No interference, but 
with a possible contribution from isovector exchange. In fit (c) the sum of 
the couplings $G_{_{ \reg \reg  \pom}}(0)$+$G_{_{ \reg \reg  \reg}}(0)$ is
allowed to be larger in the H1 data than in the fixed target data by the 
factor ${\cal R}$. The couplings shown pertain to the fixed target data.}
\label{fits}
\end{table}

\begin{figure}[h]
 \begin{center}
  \epsfig{file=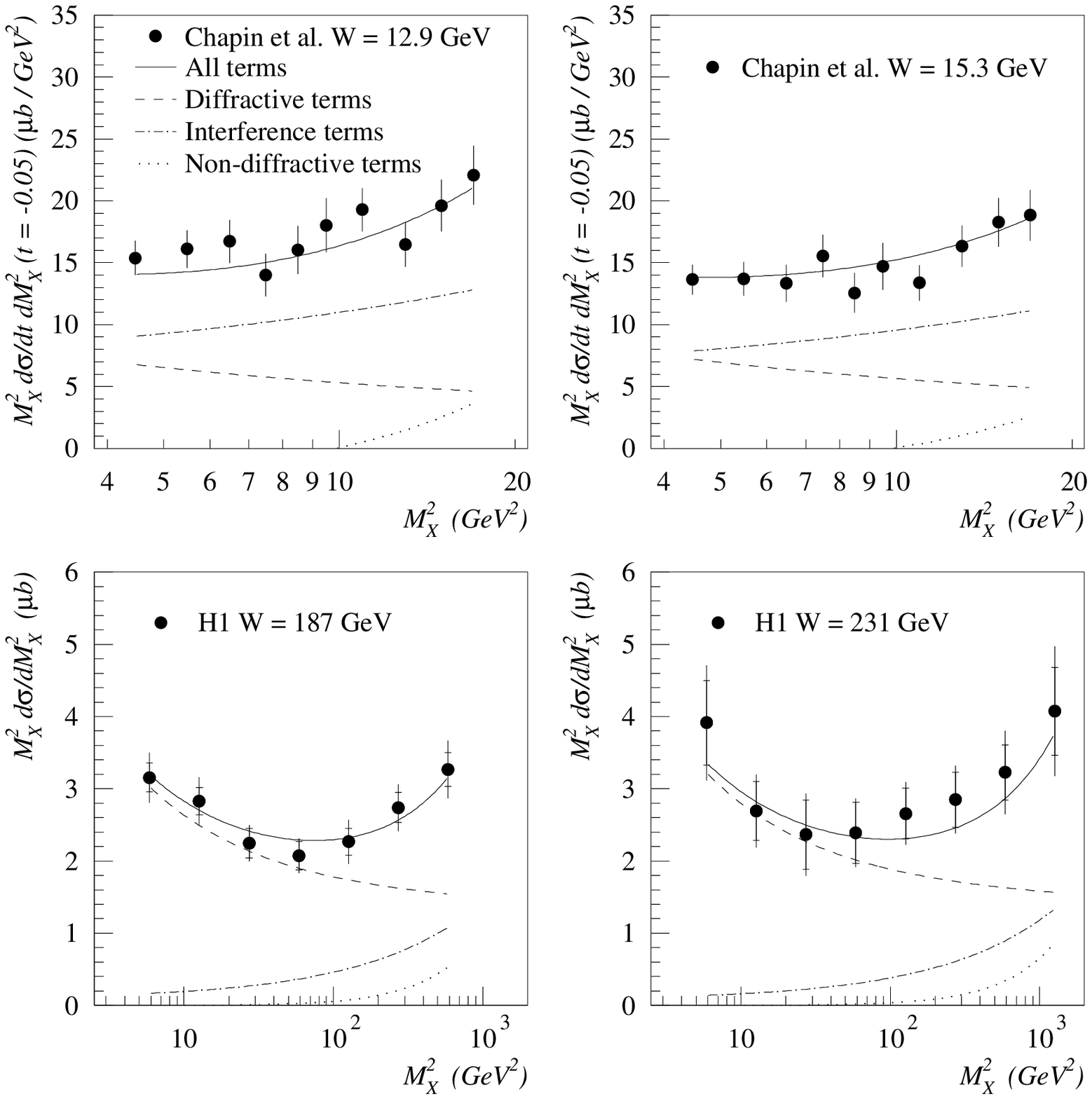,width=0.78\textwidth} 
 \end{center}
 \vspace{-0.85cm}
 \scaption  {Measurements of the quantity
$\mx^2 \, {\rm d} \sigma_{\gamma p \rightarrow XY} / {\rm d} \mx^2$ with
$\my < 1.6 \ {\rm GeV}$ and $|t| < 1.0 \ {\rm GeV^2}$ by H1 and of
$\mx^2 \, {\rm d} \sigma_{\gamma p \rightarrow Xp} / {\rm d} 
\mx^2 \, {\rm d} t$
from\protect\cite{E612:gp}\protect. For the
H1 data the inner error bars are statistical and the outer error bars show
statistical and systematic errors added in quadrature.
Overall scale uncertainties of 
13\% at $\av{\Wgp} = 12.9 \ {\rm GeV}$ and $\av{\Wgp} = 15.3 \ {\rm GeV}$,
5.2\% at $\av{\Wgp} = 187 \ {\rm GeV}$ and 6.9\% at 
$\av{\Wgp} = 231 \ {\rm GeV}$ are omitted from the errors. The
triple-Regge fit (b) with maximal constructive interference and the resulting
decomposition of the cross section is superimposed.}
  \label{gdfit}
\end{figure}

Three scenarios are examined in which subleading as well as diffractive
contributions are considered.
In fit (a) it is assumed that there is no interference between the
diffractive and non-diffractive exchanges and that there is no contribution
from isovector reggeons. Only the first four
generic terms in table~\ref{trterms} are included and the $\reg \reg \pom$
and $\reg \reg \reg$ terms have the same normalisation at all $\Wgp^2$ values.
Such a situation would be
expected to occur if the $\omega$ were the dominant subleading reggeon.
The fit parameters are shown in table~\ref{fits}a.
All experimental uncertainties are combined to determine the quoted
systematic errors. The model dependence errors arise from the
uncertainties in the parameters listed in table~\ref{params} and from
the uncertainty in the correction from $\my < 1.6 \ {\rm GeV}$.
All systematic and model dependence uncertainties are estimated from the
differences between the main fit results and those obtained with
the appropriate shifts imposed.

Interference is introduced in fit (b) by including the $\{ \pom \reg \} \pom$
and $\{ \pom \reg \} \reg$ terms.
Following~\cite{einhorn:duality,kazarinov,hidaka} two-component duality is
assumed in the reggeon-photon scattering amplitude, such that the diagonal and
non-diagonal terms are separately related for $k = \pom$ and $k = \reg$.
Maximal coherence is assumed and, through equation (\ref{schwarz}), 
the interference couplings at $t=0$ are
parameterised as $G_{_{\{ \pom \reg \} \pom}}(0) = 2 \sqrt{ \left| G_{_{\pom
      \pom \pom}}(0) \ G_{_{\reg \reg \pom}}(0) \right|}$ and $G_{_{\{ \pom
    \reg \} \reg}}(0) = 2 \sqrt{ \left| G_{_{\pom \pom \reg}}(0) \ G_{_{\reg
      \reg \reg}}(0) \right|}$. This represents the scenario in which the $f$
is the dominant subleading exchange and the $f$ and the pomeron couple 
similarly to the proton~\cite{dl:fdom,inami:roberts}. The results are presented
in table~\ref{fits}b and figure~\ref{gdfit}.  

In fit (c) effects arising from isovector exchanges are investigated. The
interference terms are not included, but the non-diffractive terms ($\reg
\reg \reg$ and $\reg \reg \pom$) are allowed to be different in the H1 and the
fixed target data. This accounts for possible additional contributions to the
H1 data from the specific terms in table~\ref{trterms} that are
marked with a star. A further free fit parameter
${\cal R}$ is therefore introduced, defined as the ratio of the 
sum of the couplings $G_{_{\reg \reg \pom}}(0)$+$G_{_{ \reg \reg \reg}}(0)$ 
in the H1 data to that in the fixed target data. If there were no isovector
contribution to the subleading exchanges, then it would be
expected that ${\cal R} = 1$. The presence of $\rho$ or $a$
exchanges would increase ${\cal R}$. The results of the fit are
presented in table~\ref{fits}c.  

From table~\ref{fits} it can be seen that
all three fits are acceptable. This demonstrates that the triple Regge
model gives a satisfactory description of both the $\Wgp^2$ and the $\mx^2$ 
dependence of the existing data on photon dissociation,
with the pomeron intercept in excess of unity.
The $\chi^2/{\rm ndf}$ value decreases markedly in fits (b) and (c) relative to
(a). This may suggest a preference for either a large interference or a 
large isovector contribution to the subleading terms, or else a 
mixture of the two. However, in light of the systematic and 
model dependence uncertainties, little can be said with any degree of 
certainty about the nature of the non-diffractive terms. 
Leading neutron measurements and data at different centre of mass energies 
would clarify the situation.
There are also further uncertainties; pion exchange may be
relevant in the fixed target data~\cite{field:fox}, and measurements at HERA
with tagged final state protons would be necessary to eliminate systematic
effects arising from the correction from $\my < 1.6 \ {\rm GeV}$ to the
elastic proton vertex. All of the fits give a $\reg \reg \reg$ contribution
that is consistent with zero, in agreement with previous
analyses of $pp$ and $\bar{p} p$ data \cite{field:fox,roberts:roy,hidaka,chu}.

The magnitude of the diffractive component at low $\Wgp$ is difficult to
constrain in light of the uncertainties.  In the fits presented here, the
diffractive contribution to the integrated cross section for $4.0 \ 
{\rm GeV^2} < \mx^2 < 0.05 \, \Wgp^2$ varies between 
$1.3$~${\rm \mu b}$ and $2.7$~${\rm \mu b}$ at
$\Wgp = 14.3 \ {\rm GeV}$, depending on the assumptions regarding the nature
of the subleading exchange. It is interesting to note 
that the value of $3.6 \pm
0.6$~${\rm \mu b}$, obtained from a fit with a more empirical treatment of
non-diffractive contributions performed by \cite{E612:gp}, is larger than any
of the values found in this analysis.

The diffractive contribution at HERA energies and the pomeron intercept
are significantly better constrained. 
A single value for the pomeron intercept is
obtained by taking an average of the three fits. The result is $\alphapom(0)
= 1.068 \pm 0.016 \pm 0.022 \pm 0.041$, where the errors are statistical,
systematic and model related respectively. The 
spread of values in the three different fits is included in the 
model dependence uncertainty and represents the dominant contribution to the 
quoted error. The value of $\alphapom(0)$ obtained is similar to those
extracted from hadronic cross sections that are governed by a soft
pomeron ($\alphapom(0) \simeq 1.081$~\cite{dl:stot}).

In analyses of $\bar{p} p$ dissociation cross sections extending to centre
of mass energies an order of magnitude in excess of those investigated
here~\cite{cdf:diss}, evidence was found of a need for unitarity
corrections in the form of a very shallow centre of mass energy 
dependence. The $\mx^2$ dependence was found to be described by the same 
pomeron intercept as the total cross section. The uncertainties associated
with the secondary exchanges in this analysis
prohibit firm conclusions regarding screening
effects for photoproduction. Although fit (a), in which neither isovector
exchange nor interference is considered, yields a value of $\alphapom(0)$ in
the region that might be expected if screening corrections were 
important \cite{glm:two}, the value is larger in fits (b) and (c) and
in all three fits presented here, the $\Wgp^2$ and $\mx^2$ dependences are 
well described with a single value for $\alphapom(0)$.

The diffractive cross section $\sigma^D$ for the process
$\gamma p \rightarrow X p$ with $4 \ {\rm GeV^2} < \mx^2 < 0.05 \, \Wgp^2$ 
is obtained at $\av{\Wgp} = 187 \ {\rm GeV}$ by integrating
the diffractive terms ($\pom \pom \pom$ and $\pom \pom \reg$) in
equation (\ref{trbasic}), with the values for the coupling constants and
pomeron intercept taken from the fits listed in table~\ref{fits}. The
results of the three fits are averaged to obtain
\begin{eqnarray}
  \sigma^D (4 \ {\rm GeV^2} < \mx^2 < 0.05 \, \Wgp^2) = 12.4 \pm 0.6 \ 
({\rm stat.}) \pm 1.4 \ ({\rm syst.}) \pm 1.7 \ ({\rm model})
  \ {\rm \mu b} \ .
  \label{sigtothi}
\end{eqnarray}
Since there is negligible non-diffractive contribution to the cross section
in the region $\mx^2<4~{\rm GeV^2}$, 
which is omitted in the fits, the corresponding measured
cross section is added to that in equation (\ref{sigtothi}) to obtain the
integrated cross section for the sum of elastic and single photon
dissociation processes. The correction from 
$\my < 1.6 \ {\rm GeV}$ to the fully
elastic case is again obtained from an average of the two Monte Carlo
simulations after tuning to match the data.  A correction factor of $0.91
\pm 0.06$ is applied in the second $\mx$ interval and $0.96 \pm 0.04$ in
the first. The latter figure is smaller than the correction used for the
remaining $\mx$ intervals because of the fast fall-off with increasing
$\my$ in the low $\mx$ region, apparent in figure~\ref{logplot} and
discussed in section~\ref{ddsec}. A correction of $1.4 \pm 1.4 \ {\rm \mu b}$,
estimated from the di-pion line shape~\cite{h1:rho}, is
added to account for the component of the cross section with $\mx < 0.4 \ 
{\rm GeV}$, which is not measured. The integrated diffractive proton-elastic
cross section is found to be
\begin{eqnarray}
  \sigma^D ( \mx^2 < 0.05 \, \Wgp^2) =
34.9 \pm 0.9 \ ({\rm stat.}) \pm 2.3 \ ({\rm syst.}) \pm 2.7 \ ({\rm model}) 
 \ {\rm \mu b} \ .
  \label{sigtotall}
\end{eqnarray}
This result is consistent, within errors, with the previous H1
measurement~\cite{h1:stot}. The \nolinebreak diffract\-ive
contribution from the process $\gamma p \rightarrow Xp$ to 
the total photoproduction cross section for \linebreak
$\mx^2 < 0.05 \, \Wgp^2$ and $\av{\Wgp} = 187 \ {\rm GeV}$ is
$22.2 \pm 0.6 \ ({\rm stat.}) \pm 2.6 \ ({\rm
  syst.}) \ \pm 1.7 \ ({\rm model}) \, \%$.

\subsection{The proton-dissociation cross section}
\label{ddsec}

No previous data or firm theoretical predictions exist that can easily be
related to the measurement with $1.6 < \my < 15.0 \ {\rm GeV}$.  When both
$\mx$ and $\my$ are large a triple-Regge analysis is in principle
possible~\cite{kaidalov:review,alberi:goggi}, though even with only two
trajectories there are a total of twelve terms with distinct dependences on
$\mx^2$, $\my^2$ and $\Wgp^2$. 
The measured double dissociation cross section is
integrated over a large region in $\my$. A full
decomposition into diffractive and non-diffractive components is not
practical without additional assumptions as to the ill-constrained behaviour
of the cross section in the low $\my$ region, the relative coupling strengths
of various reggeons to the photon and the proton and the $t$ dependence of
triple-Regge couplings.  However, as discussed below, certain features of
the large $\my$ measurements made here are as might be expected from the 
conclusions obtained in the more detailed analysis of the
proton-elastic cross section.

From the measurements for $\av{\Wgp} = 187 \ {\rm GeV}$ with 
$\my < 1.6 \ {\rm GeV}$ it is found that the diffractive
cross section in the elastic region is significantly larger than that
integrated over larger values of $\mx$ (compare equations (\ref{sigtothi})
and (\ref{sigtotall}) ). A similar pattern is observed in the dependence on
$\my$ at low $\mx$.
The cross section integrated over the range \linebreak
$0.4 < \mx < 1.26 \ {\rm GeV}$ and $\my < 1.6 \ {\rm GeV}$ is around
3.5 times larger than that for the same $\mx$ region and $1.6 < \my < 15.0 \ 
{\rm GeV}$. This clearly demonstrates the suppression of the cross section
for light vector meson production when the proton dissociates compared to that
when it remains intact. A similar behaviour is
observed in the ratios of cross sections for the processes $\bar{p} p
\rightarrow \bar{p} p$ and $\bar{p} p \rightarrow \bar{p}
Y$~\cite{cdf:diss,cdf:elas}.

Non-diffractive exchanges are likely to become more important with
increasing $\my$ as well as with increasing $\mx$. This assertion
is supported by the observation that the cross section for 
$1.6 < \my < 15.0 \ {\rm GeV}$ falls less steeply
with increasing $\mx$ than that for $\my < 1.6 \ {\rm GeV}$ 
(see figure~\ref{logplot}). In contrast to
the vector meson region, the differential cross sections in the two ranges of
$\my$ are similar in magnitude for $\mx \simeq 5 \ {\rm GeV}$.  With the
present experimental accuracy the large $\my$ cross section is well
described in the measured region by a power law of the form $\frac{{\rm d}
  \sigma}{{\rm d} \mx^2} \propto \left( \frac{1}{\mx^2} \right)^n$. In fits to
this expression for $4.0 < \mx^2 < 86.2 \ {\rm GeV^2}$, a value 
$n = 0.84 \pm 0.06 \ {\rm (stat.)} \pm 0.18 \ {\rm (syst.)}$ is obtained at
$\av{\Wgp} = 187 \ {\rm GeV}$ and 
$n = 0.85 \pm 0.14 \ {\rm (stat.)} \pm 0.18 \ {\rm (syst.)}$ at
$\av{\Wgp} = 231 \ {\rm GeV}$.

%% ========================== Conclusion ==========================
\section{Summary}

A measurement of the cross section ${\rm d} \sigma / {\rm d} \mx^2$ for the
inclusive process $\gamma p \rightarrow XY$ has been made at a centre of mass
energy more than an order of magnitude larger than previously.  A clear peak
in the cross section is observed at the lowest values of $\mx$ and $\my$,
corresponding to the elastic reaction $\gamma p \rightarrow Vp$. The cross
section falls rapidly in this region as either $\mx$ or $\my$ is increased.
At larger values of $\mx$ an approximate dependence 
${\rm d} \sigma / {\rm d} \mx^2 \sim 1 / \mx^2$ is observed, both where $Y$ is
dominantly a single proton and where it has larger masses, though the
fall in the cross section is steeper when $\my$ is small.

For the data with small $\my$ and moderately large $\mx$, a triple-Regge
decomposition of the cross section has been presented.  Despite the
simplifications necessary, it has been shown that triple-Regge
parameterisations provide a good description of both the $\Wgp^2$ and the
$\mx^2$ dependence of available photoproduction data, with parameters
constrained by other processes to which Regge phenomenology is applicable.
The data fitted do not lie in the asymptotic regime in which only the pomeron
need be considered.  An important feature of the model is that it provides a
self-consistent treatment of both diffractive and non-diffractive
contributions.

In various model scenarios investigated, the dominant exchange at $\Wgp
\simeq 200\ {\rm GeV}$ and small $\mx$ is found to be diffractive. At the
largest dissociation masses accessed, the need for a non-diffractive
contribution is apparent, which may be described with an effective trajectory
close to $\alphareg(t) = 0.55 + 0.90 t$. More measurements at different
centre of mass energies or particle identification in the system $Y$ are
required if the precise nature of the subleading exchanges is to be
determined.

The pomeron intercept is found to be in
good agreement with values extracted from total and elastic hadronic cross
sections.  In global fits to H1 and fixed target data, a value
$\alpha_{_{\pom}}(0) = 1.068 \pm 0.016 \ {\rm (stat.)} \ \pm
0.022 \ {\rm (syst.)} \ \pm 0.041 \ {\rm (model)}$ is obtained. The 
`model dependence' error arises predominantly from the spread of values 
when different scenarios are \linebreak considered for the subleading
terms. The cross section for diffractive processes of the type \linebreak 
$\gamma p \rightarrow Xp$ for $\mx^2 < 0.05 \, \Wgp^2$ is found to be 
$34.9 \pm 0.9 \ ({\rm stat.}) \pm 2.3 \ ({\rm syst.}) \pm 2.7 \ ({\rm model}) 
\ {\rm \mu b}$ at $\av{\Wgp} = 187 \ {\rm GeV}$, representing $22.2
\pm 0.6 \ {\rm (stat.)} \ \pm 2.6 \ {\rm (syst.)} \ \pm 1.7 \ {\rm (model)} \,
\% $ of the total photoproduction cross section.

The double dissociation cross section in the range $1.6 < \my < 15.0 \ {\rm
  GeV}$ and \linebreak $4.0 < \mx^2 < 86.2 \ {\rm GeV^2}$ for
$164 < \Wgp < 251 \ {\rm GeV}$ behaves approximately as ${\rm d} \sigma
/ {\rm d} \mx^2 \sim \left( 1 / \mx^2 \right)^n$. For
$\av{\Wgp} = 187 \ {\rm GeV}$ a fit to this form yields
$n = 0.84 \pm 0.06 \ {\rm (stat.)} \pm 0.18 \ {\rm (syst.)}$. The
shallowing of this dependence relative to that for the proton-elastic case
may be related to large non-diffractive contributions.

%% ========================== Acknowledgements ==========================
\section*{Acknowledgements}
We have enjoyed the benefit of several illuminating discussions with
E.~Levin and of fruitful correspondence with P.~Landshoff. We have also 
received valuable comments from E.~Gotsman,
A.~Kaidalov, U.~Maor and M.~Strikman.
We are grateful to the HERA machine group whose outstanding efforts make this
experiment possible. We appreciate the hard work of the engineers and
technicians who constructed and now maintain the H1 detector.  We thank our
funding agencies for their financial support and the DESY directorate for the 
hospitality that they extend to the non-DESY members of the collaboration.

%% ========================== Bibliography ==========================

\end{document}